# Navigating LLM Ethics: Advancements, Challenges, and Future Directions

**Junfeng Jiao [1] Saleh Afroogh*[2] Yiming Xu [3] Connor Phillips [4]**

1. Urban Information Lab, The School of Architecture, The University of Texas at Austin, Austin, TX 78712,United States. jjiao@austin.utexas.edu
2. Urban Information Lab, The School of Architecture, The University of Texas at Austin, Austin, TX 78712, United States. Saleh.afroogh@utexas.edu
3. Urban Information Lab, The School of Architecture, The University of Texas at Austin, Austin, TX 78712, United States. yiming.xu@utexas.edu
4. Urban Information Lab, The School of Architecture, The University of Texas at Austin, Austin, TX 78712, United States. connorphillips@utexas.edu

* Corresponding author: saleh.afroogh@utexas.edu

Abstract

This study addresses ethical issues surrounding Large Language Models (LLMs) within the field of artificial intelligence. It explores the common ethical challenges posed by both LLMs and other AI systems, such as privacy and fairness, as well as ethical challenges uniquely arising from LLMs. It highlights challenges such as hallucination, verifiable accountability, and decoding censorship complexity, which are unique to LLMs and distinct from those encountered in traditional AI systems. The study underscores the need to tackle these complexities to ensure accountability, reduce biases, and enhance transparency in the influential role that LLMs play in shaping information dissemination. It proposes mitigation strategies and future directions for LLM ethics, advocating for interdisciplinary collaboration. It recommends ethical frameworks tailored to specific domains and dynamic auditing systems adapted to diverse contexts. This roadmap aims to guide responsible development and integration of LLMs, envisioning a future where ethical considerations govern AI advancements in society.

**Key words**: Large Language Models, LLMs , LLM Ethics, accountable LLM, responsible LLM

# I. Introduction

Development of language models based on artificial intelligence (AI) resulted in dramatic advancements in human-computer interaction, and the nature of information dissemination and communication went through a remarkable transformation. At the forefront of these models are LLMs, with their astounding capabilities in generating texts and decoding languages across diverse domains. However, with their remarkable power comes a critical responsibility—a responsibility underscored by necessary ethical considerations that accompany their development and deployment.

The power of words holds immense responsibility, especially where the impact can directly influence individuals, families, and communities. Authors bear the weight of accountability for the spoken and written words they craft, understanding the potential repercussions they may have on those affected. In an era where technology burgeons, LLMs driven by artificial intelligence have emerged as potent tools for generating text, for example content relevant to health research and information and disinformation dissemination. These LLMs, such as ChatGPT, Jasper, Surfer, and others, wield the potential to transform the way information is conveyed and shared. However, with this transformative capability comes a heightened need for accountability and ethical consideration, as acknowledged by developers who concede the propensity for inaccuracy within AI-generated content.[1]

This paper explores the multifaceted landscape of ethical dilemmas surrounding LLMs. We embark on a journey through the intricacies of these AI-driven language models, dissecting ethical quandaries that arise in their operations, shedding light on their impact on society, and presenting prospective pathways to navigate the complex ethical web they weave. Our exploration begins with an elucidation of the convergence between Conventional Language Models (CLMs) and Pre-trained Language Models (PLMs), delving into their fundamental differences in training, causality constraints, and token representation. This comparison sets the stage for understanding the ethical considerations interwoven within the architectures of these models. Furthermore, we scrutinize the inherent biases ingrained within LLMs, dissecting their origins and their impact on AI decision-making. Our account also encompasses complex ethical aspects, acknowledging the interdisciplinary nature of addressing ethical concerns in LLMs. Additionally, as we grapple with the challenges of opacity within LLMs, we advocate for dynamic audit tools tailored to these models, emphasizing continuous monitoring, explainability techniques, and adaptable frameworks capable of navigating the ever-evolving landscape of AI-driven language models.

In this study, our pursuit is twofold: firstly, to illuminate the pressing challenges that demand immediate attention, and secondly, to present strategies for their mitigation and ethical enhancement, paving the way for responsible development and deployment of LLMs in society. Our exploration progresses in the following sequence (see, Table 1): it commences by conceptualizing LLM and ethical frameworks. The section labeled "3. Methodology" delineates the systematic review methods applied to analyze studies concerning the ethics of LLM. "4. Findings" showcases the discoveries and outcomes pertaining to primary principles and significant codes, along with their discussions in literature, comprising 13 subsections. "5. Discussion" critically examines the principal codes, fundamental values, and ethical considerations associated with LLM, along with potential strategies to address ethical concerns, thereby facilitating the responsible advancement and integration of LLMs in society. This section encompasses 11 subsections. Furthermore, the concluding thoughts and prospects concerning LLM ethics are deliberated in section 6.

**Table 1**: A road map of this study

| Section Number | Section Title | Subsection themes | | |
|---|---|---|---|---|
| 1 | **Introduction** | | | |
| 2 | **Conceptualization and frameworks** | Understanding Large Language Models | | |
| | | Ethical Theories and Approaches | | |
| 3 | **Methodology** | | | |
| 4 | **Findings** | 4.1. | Ethical Concerns in LLM | |
| | | 4.2. | Ethical Frameworks and platforms in LLM | |
| | | 4.3. | Bias and Fairness in LLM | 4.3.1. Different types of bias in LLM |
| | | | | 4.3.2. Fairness in LLM |
| | | | | 4.3.3. Bias and Fairness Detection Methods |
| | | 4.4. | Privacy and Data Security in LLM | |
| | | 4.5. | Misinformation and Disinformation in LLM | |
| | | 4.6. | Accountability and Governance in LLM | |
| | | 4.7. | Case Studies in LLM ethics | |
| | | 4.8. | Mitigation Strategies in LLM ethics | |
| | | 4.9. | Transparency in LLM | |
| | | 4.10. | Censorship in LLM | |
| | | 4.11. | Intellectual Property and Plagiarism in LLM | |
| | | 4.12. | Abusive LLM, hate speech and cyber-bullying | |
| | | 4.13. | Auditing LLM | |
| | **5. Discussion** | 5.1. | The Advantages of Pre-Trained Models in Integrating Normative Ethics in LLM | |
| | | 5.2. | Embracing Multidisciplinary Perspectives Beyond Engineering for Ethical AI in LLMs | |
| | | 5.3. | Protecting Privacy in Language Models amidst Growing Data Concerns | |
| | | 5.4. | Ethical Uniqueness: Specialized Considerations for LLMs | |
| | | 5.5. | Hallucination and Distorted Realities in LLMs | |
| | | 5.6. | Verifiable Accountability and Citation Integrity System in LLMs | |
| | | 5.7. | Diversifying Case Studies in LLMs | |
| | | 5.8. | Decoding Censorship Complexity in LLMs | |
| | | 5.9. | Breaking the Black Box and Crafting Dynamic Audit Tools for LLMs | |
| 6 | **Conclusion and Future directions** | | | |

## II. Conceptualization and frameworks

Large language models (LLMs) represent a transformative advancement in artificial intelligence, enabling sophisticated natural language processing and generation. This section explores their foundational principles and ethical implications. Section 2.1 examines the technical underpinnings of LLMs, including their predictive mechanisms and societal applications. Section 2.2 discusses key ethical theories and approaches guiding their responsible development and deployment.

## 2.1. Understanding Large Language Models

Language models (LMs) are essential tools within the scope of natural language processing (NLP) for understanding and predicting probability distributions within sequences of linguistic units, such as words or phrases. Their primary function is to anticipate the likelihood of tokens (textual units, often broken down into sub-word units) occurring within a given text sequence. These models, particularly the generative ones, function in an autoregressive manner, predicting the probability distribution of a token based on its preceding tokens. The predictive mechanism of LMs involves using the chain rule of probability and conditional probabilities to estimate this joint probability:

$$P(u1, u2, \cdots, ut) = P(u1)P(u2|u1)P(u3|u1, u2) \cdots P(ut|u1, ...ut-1),$$

wherein 'u' symbolizes a sequence comprising T tokens, and P(u1) signifies the probability of the first unite u1, P (u2 | u1) signifies the probability of the u2 given u1, representing the probability of u2 occurring after u1, and so on up to ut.[2]

A LLM significantly augments the size of an LM, incorporating a large number of model parameters, typically ranging from tens of millions to billions [3]. These models have exhibited some unexpected abilities and emerging skills. [4] [5] To enhance its proficiency, an LLM undergoes extensive training using large volumes of diverse data. This augmented size and extensive training enable LLMs to grasp intricate linguistic patterns, thereby improving their ability to generate more coherent and contextually appropriate text. The methodological foundation of LLMs involves leveraging immense computational power coupled with sophisticated algorithms to process and analyze colossal amounts of textual data. This enables these models to acquire a deep understanding of language structures and semantics, thereby generating a text that closely resembles human-written content.

LLMs represent a milestone in the field of artificial intelligence, embodying sophisticated neural architectures designed to comprehend and generate human-like text. The historical trajectory of these models indicates a progressive evolution from early language processing frameworks to the advent of contemporary behemoths like Google Gemini, Claude, GPT, DeepSeek, Meta Llama, Grok 3. With billions of parameters, these models demonstrate excel in natural language understanding, generation, and translation.[3] Their key attributes include contextual understanding, semantic coherence, and adaptability, enabling them to perform diverse language-related tasks with remarkable finesse. Furthermore, they continuously improve through their capacity for unsupervised learning and adaptation to varying linguistic contexts.

LLMs have applications across multifaceted domains, highlighting their significance and pervasive impact on modern society. These models expedite breakthroughs in industries like healthcare, education, and finance by facilitating language translation, text summarization, conversational AI, and content generation. These models can be utilized in information retrieval, sentiment analysis, and personalized content recommendation. Further, their transformative capabilities are evident in their potential to revolutionize human-computer interaction. With the progress of these models, their use in diverse applications has profound implications for reshaping the landscape of human-computer interfaces and the broader technological paradigm.[6][7]

### 2.2.Ethical Theories and Approaches

There are various ethical considerations surrounding LLMs, involving various moral frameworks such as Utilitarianism, Deontology, Virtue Ethics, and more. [8], [9] These frameworks and related theories can lay the ethical grounds for developing, implementing, and using LLMs. Some multidimensional approach are also proposed for embedding ethical concerns into LLM development, as it involves integrating ethical considerations throughout the design process, ensuring diversity representation in data collection, fostering transparency in model development, and implementing mechanisms for accountability and ongoing evaluation of the model's outcomes. Furthermore, it requires interdisciplinary collaborations where ethicists, technologists, policymakers, and stakeholders engage to establish comprehensive frameworks that align ethical principles with practical applications in the realm of large language models .[10]

## III. Methodology

We conducted an inclusive and systematic review of academic papers, reports, case studies, and frameworks regarding LLM ethics, written in English. Given that there is not a specific database on LLM ethics in particular, we used the Preferred Reporting Items for Systematic Reviews and Meta-Analyses (PRISMA) framework to develop a protocol in this review (Figure 1).[11]

In order to conduct a comprehensive review of the relevant studies, we followed two approaches. First, we manually searched for the most related papers on LLM ethics: 17 papers were identified through online search after the removal of duplicate files. Secondly, we fulfilled a keyword-based search (using the Google Scholar search engine) to collect all relevant papers on the topic. This search was accomplished using the following keyword phrases: (1) "ethics + large language models" which provided 14 relevant result pages of Google Scholar, (2) "ethics + large +language+ models" for which the first six result pages were reviewed, and (3) "LLM + ethics," for which the first nine result pages of Google Scholar were reviewed.

Moreover, the following keywords "transparency/ privacy/ fairness / bias /accountability/ mitigation / misinformation / hate speech / cyber-bullying / copyright / Censorship /auditing / limitations / Case studies + large language models/LLM" were reviewed respectively in first 3/6/5/6/4/16/6/6/6/6/4/6/10 pages of google scholar and included because of their central role in the research as the major known (based on a preliminary review) ethical considerations of LLM. Additionally, the search was suspended within results for each search term due to limited appearances of new relevant papers on the following pages.

The results of the search were 456 relevant papers (which were selected based on the semantical keywords relevancy), out of 1147 (which appeared on the result pages). Afterward, the duplicated papers were eliminated from the analysis. We selected the 192 target papers for this systematic review based on the following two inclusion/exclusion criteria. First, articles that were published in academic journals were included. Second, the dominant topic of the papers (or a significant part of it) was LLM ethics. To this end, the papers' main sections were reviewed to understand their dominant topic rather than only relying on the title and papers' keywords (As a result, 264 papers were excluded.)

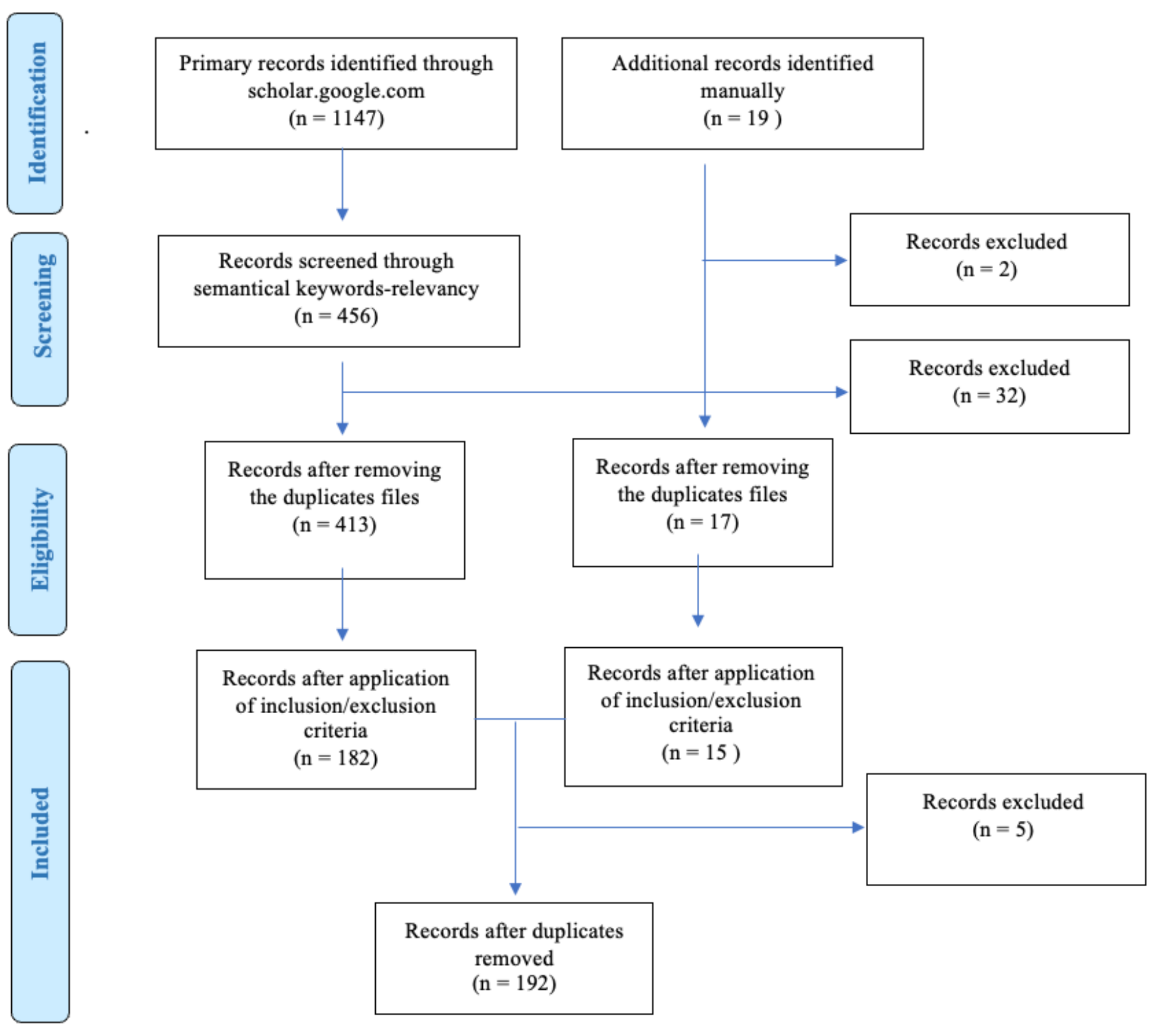

**Fig 1**. Developed PRISMA flow diagram for ethics of large language models (LLM Ethics)

## IV. Findings and Resultant Themes

The qualitative analysis on the selected papers was performed by four researchers who critically read the papers and who developed the eight major key codes as the building blocks of the categorization of the review result in the next step of this research (Table 2).

**Table 2**: Major and minor codes included in the reviewed papers.

| | Major ethical codes | Number of reviewed papers | Minor ethical codes |
|---|---|---|---|
| 1 | Ethical Concerns in LLM | 14 | Bias and fairness, Privacy and data security, Misinformation and disinformation, Transparency and accountability, Intellectual property and plagiarism, Access and inequality, Reinforcement of societal prejudices, Data breaches, Public safety, False or misleading content, Manipulation of public opinion, Discrimination, Environmental impact, Marginalized groups. |
| 2 | Ethical Frameworks and platforms in LLM | 12 | Responsible deployment, Human-centered framework, Users' mental models, Utility of use cases, User-centric measures, Deontological ethics in NLP, Risk assessments and regulations, Human autonomy in AI systems, Intentional misuse, Unintentional harm, Bottom-up approaches, Top-down framework, Common-sense morality datasets, Two-step framework, Ethics by Design, Default ethical settings, Embedded ethics. |
| 3 | Bias and Fairness in LLM | 23 | Social biases, Language biases, Representation biases, Income, Race, Lifestyle choices, Gender biases, Dialects, Linguistic styles, Social inequalities, Representation biases, Misrepresented groups, Demographic representations, Cultural representations, Socioeconomic representations, Reinforcing stereotypes, Perpetuating inequalities, FaiRLLM benchmark, Group and individual fairness, Real-world datasets, Superior fairness. |
| 4 | Privacy and Data Security in LLM | 10 | Safeguarding sensitive information, Privacy risks, Mitigation techniques, Privacy-preserving methods, Differential privacy, Maintaining utility and accuracy, Integrating privacy-preserving techniques, Rigorous data security standards. |
| 5 | Misinformation and Disinformation in LLM | 13 | Unintentional propagation, Mechanisms to detect misinformation, Misleading content, Influence on public opinion, Discourse shaping, Contextual accuracy, Contextually inappropriate content, Unintentional or intentional biases, Regulatory compliance of misinformation. |
| 6 | Accountability and Governance in LLM | 22 | Auditable decision-making, Accountability in the Healthcare sector, Risks of AI recommendations, Documentation of data origins, Interactive model cards, Integration of a citation mechanism in LLMs, Stakeholder accountability, External scrutiny methods, Red-teaming and auditing, ASPIRE framework, Knowledge Management Systems (KMS). |
| 7 | Case Studies in LLM ethics | 25 | Healthcare sector, Patient data confidentiality, Impact on research integrity, Plagiarism, Changing landscape of academic publishing, Educational sector, Ethical use of AI in student evaluations, Workplace communication automation, Ethics judgment, Streamlining and automation, Digital divide within society, Replacement of human job. |
| 8 | Mitigation Strategies in LLM ethics | 26 | Multifaceted challenge, Bias mitigation, Privacy protection, Hallucinations prevention, Social bias, Gender bias and stereotypes, Dataset enhancement, InfoEntropy Loss function, Adversarial learning, Semantic similarity task, Embedding Purification (E-PUR), Comparative testing, E-PUR method and baseline methods, Hallucinations Prevention, Output verification, Proactive detection, Participatory design. |
| 9 | Transparency in LLM | 7 | Clarity and understandability, Facilitating accountability, Interpretable models, Engaging in open practices, Chaining LLM steps, Model reporting, Publishing evaluation results, Undocumented data, Questionable legality. |
| 10 | Censorship in LLM | 1 | Benefits of censorship, Preventing harmful outputs, Censorship Concerns, Subjectivity in Censorship, Risk of infringing on free speech, Transparency in censorship policies, Beyond semantic restrictions, Allowed outputs, Implementing censorship. |
| 11 | Intellectual Property and Plagiarism in LLM | 6 | Intellectual property rights, Copyrighted material, Watermarking models, Ownership in academic research, Fair use, Liability, Highly resource-intensive. |
| 12 | Abusive LLM, hate speech and cyber-bullying | 7 | Facilitation of harmful content, Abusive language, Hate speech, Toxic or abusive content, LLMs for detection of online abusive language, Utilizing open-source pre-trained Llama 2 model, Detecting cyberbullying on social media platforms. |
| 13 | Auditing LLM | 7 | Systematic evaluation, ChatGPT-powered causal auditor, Discrete optimization, Continuous auditing tools, Frequency of audits, Hindrance to audits, Cultural and global boundaries, Dynamic and adaptable auditing practices. |

### 4.1. Ethical Concerns in LLM

LLMs like ChatGPT and LLaMA give rise to various ethical concerns with significant implications for society [12], [13], [14], [15], including bias and fairness, privacy and data security, misinformation and disinformation, transparency and accountability, intellectual property and plagiarism, access and inequality [12], [13], [14], [15], [16], [17], [18], [19], [20].

Biases may exist in data with which LLMs have been trained. These may affect their functioning, resulting in outputs that accentuate stereotypes or involve unjust discriminations against certain groups, negatively affecting marginalized groups [12], [14]. Moreover, the data with which LLMs are trained may contain personal, sensitive, or proprietary information, which raises worries about privacy and data security, risking individual and public safety and compromising trust in digital systems [12]. The misinformation and disinformation concerns are raised by the fact that LLMs are capable of generating convincing but potentially false or misleading content. This can be used to spread misinformation, manipulate public opinion, or create fraudulent materials, which has severe consequences for public health, democracy, social harmony [15], [18], [21], [22]. Transparency and accountability concerns result from the black-box nature of LLMs, which makes it difficult to determine responsibility for harmful outputs or decisions [12]. The ability of LLMs to generate texts that closely resemble human writing raises concerns about intellectual property and plagiarism [13]. In addition, the benefits of LLMs might be unequally distributed, exacerbating existing inequalities, raising concerns on access and inequality [12], [23].

Existing studies have explored these ethical concerns in LLM using surveys, experiments, and reviews. [12] examines the ethical aspects of LLMs like ChatGPT, using established methods for technology ethics analysis. It identifies benefits and challenges, including issues of social justice, autonomy, safety, bias, accountability, and environmental impact. Moreover, the study highlights the need for a broader approach to AI ethics, focusing on stakeholder engagement and holistic policy interventions. [15] summarized the risks from LLMs into six areas and reviewed 21 risks in detail, covering discrimination, data security, misinformation, environment issues, and so on. [16] investigated the effects of biases on LLMs to understand the practical aspects of implementing LLMs. Some studies focused on the ethical concerns in specific scenarios. For example, [13] analyzed ChatGPT's ability to offer advice on cheating in assessments by a series of 'conversations' held with multiple instances of the model. [21] explored the ability of current LLMs to blend advertisements with organic search results, indicating that their proficiency in merging ads with relevant topics is clear.

As discussed in literature, the ethical concerns in LLMs arise due to factors inherent in their design, training, and potential applications [12], [13], [15], [16], [21], [23], [24], [25]. Dealing with these concerns involves a multi-faceted approach that includes technical strategies (e.g., algorithms to detect and reduce biases), policy development (e.g., privacy protection regulation), ethical guidelines (e.g., respecting intellectual property), and stakeholder engagement (e.g., collaboration between industry, academia, and regulatory bodies).

### 4.2. Ethical Frameworks and platforms in LLM

The ethical frameworks surrounding LLMs are multifaceted and crucial for their responsible deployment (See Table 3). In order to mitigate potential dangers and misconceptions, a human-centered evaluation framework for LLMs is proposed. This framework places emphasis on understanding users' mental models, assessing the utility of use cases, and addressing cognitive engagement in human-AI interaction.[26]

**Table 3**: Ethical Frameworks and Platforms for LLMs

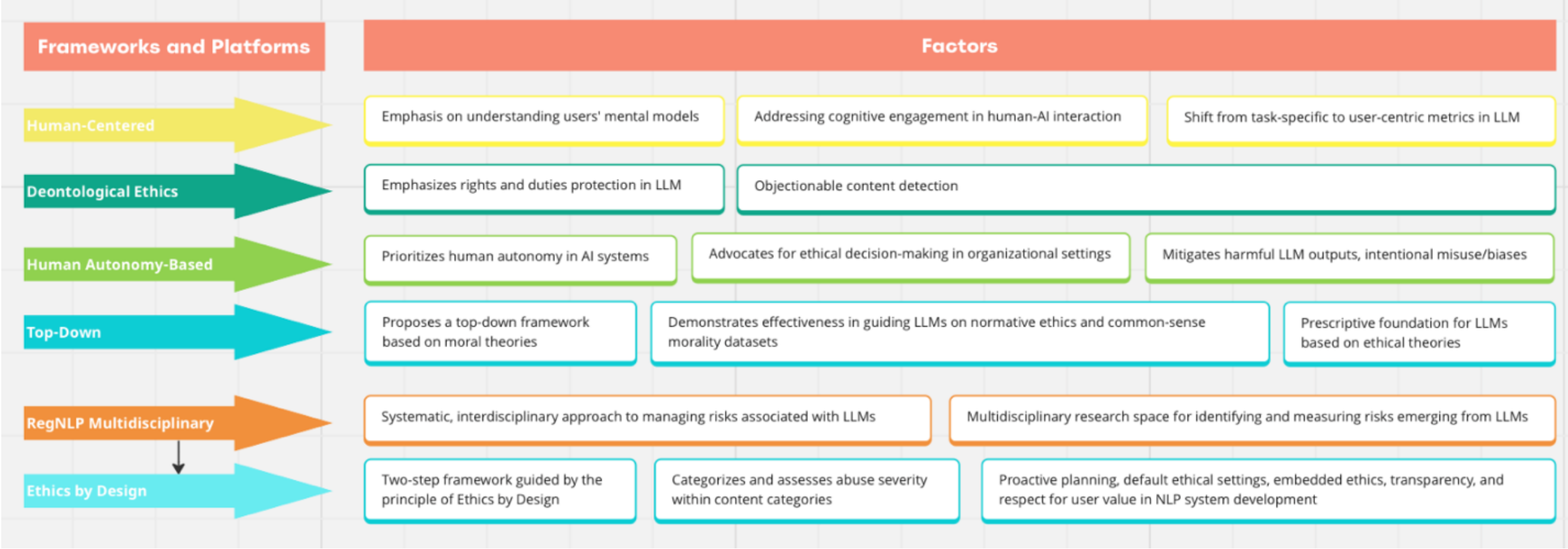

Given the need for new evaluation metrics, a shift from task-specific to user-centric measures is also proposed in employing LLMs. [27] Some scholars outlined seven trustworthiness categories and 29 sub-categories related to reliability, safety, fairness, explainability, adherence to norms, and robustness. [28] Moreover, [29] provides a solid foundation for an ethical framework for LLMs by invoking deontological ethics in Natural Language Processing (NLP). The paper highlights the application of this framework in question-answering systems and objectionable content detection through four case studies, emphasizing the robustness of deontological ethics in safeguarding individuals' rights and duties. Regarding chatbots in healthcare, particularly in hard-to-reach communities, some scholars propose a guide for evaluating and monitoring chatbot technology deployment, recommending the incorporation of medical ethics into AI regulatory frameworks in order to enhance risk assessments and regulations.[30]

[31] underscores the importance of prioritizing human autonomy in AI systems, addressing concerns about intentional misuse, unintentional harm caused by biases, and societal impacts. It advocates for ethical decision-making in organizational settings and curated datasets to mitigate harmful language model outputs. Some also critiques bottom-up approaches and proposes a top-down framework based on moral theories, demonstrating its effectiveness in guiding LLMs on normative ethics and common-sense morality datasets. [32] Highlighting polarization in AI debates, some scholars also argue for a systematic, interdisciplinary approach to managing the risks associated with LLMs.They propose RegNLP as a multidisciplinary research space for identifying and measuring risks emerging from LLMs and NLP technology.[33]

In their examination of ethical considerations in detecting online abusive content, some scholars raise concerns despite high accuracy. They proposes a two-step framework guided by the principle of Ethics by Design, which categorizes and assesses abuse severity within content categories.[34] Discussing ethical outcomes in NLP system development, some scholars advocate for an "Ethical by Design" approach, which promotes best practices such as proactive planning, default ethical settings, embedded ethics, transparency, and respect for user value.[35], [36].

Discussing ethical frameworks in LLMs, some studies addresses methodological challenges in developing morally informed AI systems, particularly in the case of recent models such as GPT-3 and RoBERTa. It notes their reliance on fine-tuning with specific data and the need for a prescriptive foundation based on ethical theories.[37]

## 4.3. Bias and Fairness in LLM

Large Language Models (LLMs) can inherit and amplify biases from their training data, raising ethical and operational concerns. This section examines (4.3.1) key types of biases in LLMs, (4.3.2) fairness evaluation and mitigation strategies, and (4.3.3) methods for detecting biases and assessing fairness.

### 4.3.1. Different types of bias in LLM

Language Large Models (LLMs) like ChatGPT and LLaMA can exhibit biases. These biases generally result from their training data and the algorithms used in their development. The biases in LLMs can be categorized into social biases, language biases, and representation biases.(See Table 4)

Social biases in LLMs have been extensively studied by scholars recently. Social biases reflect model biases related to various aspects of social status and characteristics, such as gender, age, income, race, occupation, education, and lifestyle choices [38], [39], [40], [41], [42], [43], [44], [45], [46], [47], [48], [49], [50], [51]. These biases are generated due to the nature that LLMs can perpetuate biases in the training data, which reinforces stereotypes and unequal performance in content generation related to different social groups. [46] conducted an examination of gender biases in reference letters generated by LLMs, focusing on biases in language style and lexical content. The results indicated substantial gender biases in the LLM-generated recommendation letters. [42] employed ChatGPT and LLaMA to create news content, using headlines from two newspapers recognized for their unbiased reporting. They assessed the gender and racial biases in the content generated by these LLMs by comparing it with the original news articles. The study found that the LLM-generated content displayed significant biases, particularly showing discrimination against women and Black individuals.

Language biases refers to the preferential treatment or unequal performance of LLMs across different languages, dialects, or linguistic styles. These biases can lead to unequal service quality or information accuracy, reinforcing social inequalities in education, communication and business [38], [45]. In addition, [38] found that biases regarding gender, race, age, religion, and social class exist when the LLMs are used for translation tasks.

**Table 4**: Major identified biases in LLMS

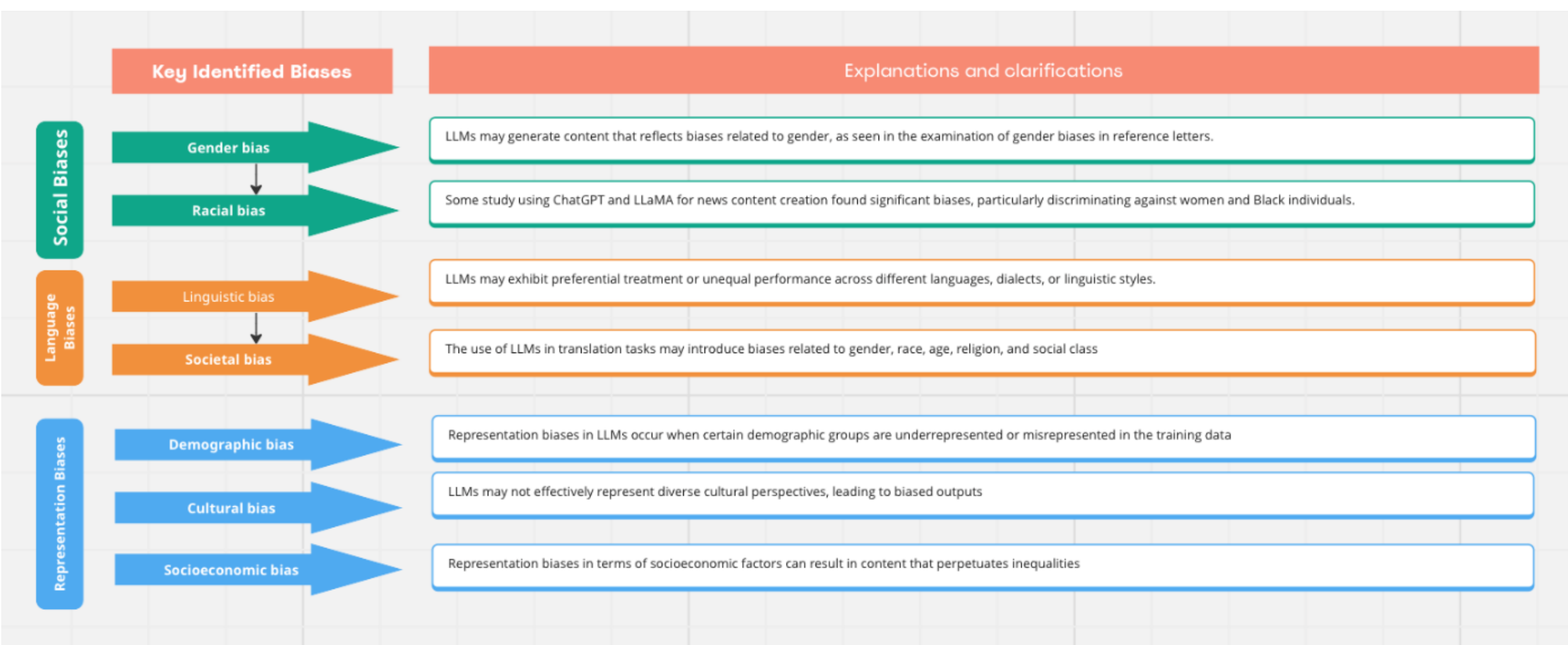

| | Key Identified Biases | Explanations and clarifications |
|---|---|---|
| Social Biases | Gender bias | LLMs may generate content that reflects biases related to gender, as seen in the examination of gender biases in reference letters. |
| | Racial bias | Some study using ChatGPT and LLaMA for news content creation found significant biases, particularly discriminating against women and Black individuals. |
| Language Biases | Linguistic bias | LLMs may exhibit preferential treatment or unequal performance across different languages, dialects, or linguistic styles. |
| | Societal bias | The use of LLMs in translation tasks may introduce biases related to gender, race, age, religion, and social class |
| Representation Biases | Demographic bias | Representation biases in LLMs occur when certain demographic groups are underrepresented or misrepresented in the training data |
| | Cultural bias | LLMs may not effectively represent diverse cultural perspectives, leading to biased outputs |
| | Socioeconomic bias | Representation biases in terms of socioeconomic factors can result in content that perpetuates inequalities |

Representation biases in LLMs occur when certain groups, perspectives, or types of information are underrepresented or misrepresented in the training data of these models, including demographic, cultural, and socioeconomic representations. The consequences of representation bias can reinforce stereotypes, perpetuating inequalities, and failing to serve diverse user groups effectively [41], [44], [45], [48].

It is worth noting that bias in language models is usually multidimensional and content-specific, extending beyond simple polarity distinctions such as positive or negative sentiment [210], [211]. It encompasses deeper issues of what content is represented, how frequently it appears, and whose perspectives are prioritized or omitted. This includes disparities in visibility across social identities, geographic regions, and topic domains, all of which can shape user perceptions and reinforce structural inequalities. Moreover, social, language, and representation biases are often deeply interconnected. For example, biases in language use, such as favoring standard dialects or dominant languages, can reflect and reinforce existing social hierarchies. Similarly, underrepresentation of certain groups in training data can simultaneously manifest as both social and representational bias. In multilingual settings, translation can propagate or even amplify the social biases present in the target language's cultural context. These overlaps underscore the complexity of bias in LLMs and highlight the need for nuanced, intersectional approaches to detection and mitigation.

While these categories of biases are interconnected, they differ in terms of how commonly they occur, how serious their implications are, and how challenging they are to detect and mitigate. Social biases are the most visibly harmful and most frequently flagged during public audits. They are detectable through bias benchmarks (e.g., StereoSet, CrowS-Pairs), but mitigation is difficult due to the scale of training data and the nuanced nature of prejudice [196],[197]. Alignment strategies like RLHF help, but may also introduce new forms of censorship or bias. Language biases are systematic, favoring well-resourced languages (especially English) and standard dialects. Tools like XNLI and FLORES help in detecting these, but low-resource language support remains poor [205], [206]. Mitigation is tough due to data scarcity, model scaling challenges, and uneven community engagement in model training. Representation biases are hardest to detect and fix. They don’t always produce offensive content but can lead to underrepresentation of certain groups or overemphasis on dominant cultural narratives. These biases often originate from skewed training datasets (e.g., overuse of Western-centric sources). Mitigation requires rebalancing or augmenting data, which is expensive, time-consuming, and complex to validate.

### 4.3.2. Fairness in LLM

Recent AI research has increasingly focused on the issue of fairness in LLMs, driven by an expanding recognition of their potential to perpetuate or even magnify existing biases. Research in this area primarily concentrates on assessing the fairness of LLMs and developing strategies to improve their fairness. [52], [53], [54], [55], [56], [57], [58], [59], [60]

Multiple scholars have proposed evaluation methods to assessing the fairness of LLMs. [54] introduced a new benchmark named FaiRLLM. This benchmark includes meticulously designed metrics and a dataset that considers eight sensitive attributes within two specific recommendation contexts: music and movies. Their evaluation of ChatGPT revealed that the model continues to show elements of unfairness in relation to some of these sensitive attributes. [53] examined ChatGPT's performance in high-stakes areas such as education, criminology, finance, and healthcare. They employed a detailed evaluation framework that accounted for both group and individual fairness and analyzed disparities in ChatGPT's responses to biased and unbiased prompts.

Researchers have been developing methods to improve the fairness in LLMs, focusing on improvements in both the model training process and the design of the prompts. [59] proposed an approach to make references to demographic groups unrelated to their portrayal in the text, thus reducing social bias. They also introduce a method to estimate this correlation's upper limit using importance sampling and a natural language corpus. Empirical tests on real-world benchmarks show that this method effectively improves fairness without compromising language modeling proficiency. [55] leverages a personalized prefix prompt and a prompt mixture designed to boost fairness in relation to various sensitive attributes. Their experiments with two real-world datasets reveal that the UP5 model demonstrates superior fairness compared with benchmark fairness-aware recommendation models.

### 4.3.3. Bias and Fairness Detection Methods

Bias and fairness detection in large language models (LLMs) is essential to ensure ethical, safe, and equitable AI systems. LLMs often inherit or amplify societal biases present in their training data. To address this, researchers have developed a range of bias and fairness detection methods, which can be grouped into several categories, including template-based bias probing, likelihood and score-based fairness metrics, embedding space analysis, and behavioral audits.

The template-based bias probing methods use controlled sentence templates to isolate and test specific biases. For example, [196] presents StereoSet, a large-scale natural dataset in English, to measures bias across four domains: gender, race, religion, and profession by evaluating if models prefer stereotypical over anti-stereotypical completions. [197] introduces the Crowdsourced Stereotype Pairs benchmark (CrowS-Pairs) that contains sentence pairs with stereotypical and anti-stereotypical forms across social dimensions (race, gender, age, etc.) to measure social bias in LLMs. Likelihood and score-based fairness metrics compare output probabilities, completions, or toxicity scores across demographic groups. For example, [198] evaluates toxicity of LLM outputs across prompts involving different identity groups. Embedding space analysis measures biases in the vector space of word embeddings by word embedding association test [199] and sentence encoder association test [200]. Behavioral audits test how LLMs behave in real-world interaction scenarios to see if the model generates biased outputs [201].

### 4.4. Privacy and Data Security in LLM

Most academic research on data security and privacy in LLMs reflects a growing recognition of the critical need to safeguard sensitive information [60][61][62], as these models present unique privacy challenges. LLMs demonstrate capacity for inadvertent memorization and reproduction of sensitive information from training data, including personal identifiable information such as names, addresses, and phone numbers, financial data including credit card numbers and bank account details, medical records and health information, and private communications and personal documents. Research has demonstrated that LLMs can be prompted to reveal memorized training data through targeted extraction attacks [60], posing significant risks for individuals, whose personal information was included in training datasets without consent.

Inference-time privacy concerns emerge when users interact with LLMs. User queries frequently contain sensitive personal information, conversation histories can reveal behavioral patterns and preferences, and model responses may inadvertently disclose information about other users. Third-party integrations expand this vulnerability, creating complex chains of data exposure that are difficult to monitor and control. These privacy risks intersect with regulatory frameworks such as GDPR and CCPA, creating compliance challenges for LLM deployment. Right to deletion requirements conflict with the technical difficulty of "untraining" specific data, data minimization principles clash with LLMs' need for vast training datasets, and consent requirements prove practically impossible to implement retroactively for existing models.

Researchers are actively exploring innovative techniques to mitigate privacy risks inherent in LLMs [63][64], with key areas including developing privacy-preserving methods like differential privacy [65], discriminator use [66], and matrix-vector functions [67]. These approaches aim to protect user data while maintaining utility and accuracy of LLMs. Research directions increasingly focus on integrating privacy-preserving techniques directly into architecture and training processes of LLMs [68], [69], seeking balance between advanced language processing capabilities and data security standards.

### 4.5. Misinformation and Disinformation in LLM

Many scholars are concerned with how LLMs can unintentionally propagate or amplify false information [70], given their capacity to generate realistic and coherent text. The distinction between misinformation, defined as inadvertently false information, and disinformation, characterized as deliberately false information, becomes crucial when examining LLM-generated content. LLMs can unintentionally propagate or amplify false information through several identified pathways, including training data contamination with existing false information, hallucination phenomena generating plausible but incorrect facts, outdated information presented as current due to training data temporal limitations, and accurate information being presented inappropriately for specific circumstances.

More concerning is the demonstrated potential for deliberate misuse of LLMs in disinformation campaigns. Research has documented LLM capabilities for generating convincing fake news articles at scale, creating misleading social media content, automating generation of conspiracy theories and propaganda, and enabling sophisticated manipulation of public opinion through targeted messaging. Scholars are developing mechanisms to detect and mitigate misinformation spread by LLMs [71]–[74], focusing on automated fact-checking systems that verify LLM outputs against reliable sources, tracking to identify AI-generated content, real-time monitoring systems for detecting misinformation spread, and training methods to make LLMs more discerning in processing and relaying factual information [77]–[79].

Key research areas include understanding the models' susceptibility to biases that might lead to the generation of misleading content [75], [76] and designing training methods to make LLMs more discerning. The societal impact of LLM-generated misinformation extends to fundamental questions about information reliability and democratic discourse. Studies examine LLMs' potential role in influencing public opinion and shaping discourse [80], [81], while the challenge of equipping LLMs with capabilities to evaluate truthfulness of input data and contextual accuracy of generated information [82] remains unsolved. Social media platforms and search engines are implementing detection systems for AI-generated content, while governments consider regulations requiring disclosure of AI-generated information. However, the rapid advancement of LLM capabilities often outpaces detection technologies, creating ongoing challenges for information integrity in digital environments.

### 4.6. Accountability and Governance in LLM

Accountability is a pivotal principle essential for ensuring transparent and auditable decision-making processes. It serves as a cornerstone in bolstering public trust and meeting the increasing expectations set by regulatory bodies [84]. Large Language Models may amplify and solidify pre-existing biases ingrained within their training data. This tends to produce misleading or inaccurate information. Consequently, ensuring accountability is of utmost importance when incorporating these models into the Healthcare sector [85]. Healthcare professionals have expressed concerns about added tasks managing technology, the potential risks of unreliable AI recommendations, and the potential loss of human connection in patient care due to AI integration.[1]

In addition to LLMs, the issue of generating misleading information also applies to Large Vision Language Models (VLMs). As confirmed by [86], objects with a high frequency of occurrence or those that co-occur with elements in visual data are particularly susceptible to misinterpretation by VLMs. The creation of counterfactual speech further complicates the pursuit of accountable AI. For example, fabricating references in scientific article composition [87] and inventing fictitious legal cases within the legal domain [88] underscore the inherent risks associated with the utilization of ChatGPT in critical domains. Choudhury et al. [89] underscore the importance of addressing accountability in healthcare by educating practitioners on how to evaluate AI recommendations, acknowledging challenges posed by heavy workloads and limited statistical training. To effectively tackle these hurdles, the paper suggests implementing policies that establish accountability measures for both clinicians and AI systems, highlighting the importance of education and regulatory frameworks in the healthcare sector. [85]

The call for greater transparency and accountability in high-stakes AI research requires a fundamental overhaul of incentives, turning away from rapid advancement that may overlook meticulous scientific inquiry [90]. This transformation involves acknowledging the critical role of data work, incentivizing thorough documentation of data origins, and addressing the need to identify and mitigate any adverse effects [91]. By leveraging established frameworks from software engineering and infrastructure, fields such as AI and NLP can adopt robust structures to enforce dataset accountability [92]. Furthermore, integrating human-centric approaches like interactive model cards [93] holds promise in promoting comprehensive documentation and accountability within these domains.[94] Moreover, open-source initiatives in conversational text generators play a critical role in fostering accountability by evaluating and highlighting various degrees of openness across key dimensions such as data legality, documentation standards, and accessibility. This impacts fairness and accountability from data collection to model deployment [95]

Research suggests that the scientific language used by AI, such as ChatGPT, might mislead users about its reliability, which highlights the need for accessible references generated by humans [96][97][98]. Scholars emphasize the significance of accountability in LLMs by proposing the integration of a citation mechanism to address issues related to intellectual property, ethics, transparency, and verifiability. By adopting a citation system that attributes sources, these models can foster accountability, respect intellectual property rights, and uphold information integrity, thereby contributing to the development of more responsible and trustworthy AI systems[99][100]; Furthermore, certain frameworks are proposed for dataset accountability [101]. Cacciamani *et al.* [102] proposed CANGARU Guidelines for data scientists, which encompass ethical accountability, reproducibility, disclosure, and proper reporting concerning GAI/GPTs/LLMs in academic papers, aiming to promote critical adherence among academics, authors, editors, reviewers, publishers, and readers.

Some scholars stress the necessity of holding stakeholders accountable in the development and use of LLMs through external scrutiny methods like red-teaming and auditing, guided by the ASPIRE framework. Anderljung [103] examines the implementation of an AI system at BlockScience, an engineering company, where an LLM was integrated into their internal Knowledge Management System (KMS). This study delves into the challenges and implications of this integration, highlighting the importance of understanding accountability in human-AI interactions within organizational contexts and the need for strategies aligning AI technologies with human interests.[104]

### 4.7. Case Studies in LLM ethics

Ethical case studies of LLMs across various sectors reveal distinct patterns of challenges and responses. Healthcare applications demonstrate some of the highest consequence severity for ethical failures, with research examining implications of LLMs in patient data confidentiality and consent [104], [105], medical training and education [106], [107], and potential biases in medical recommendations [108], [109]. Patient data confidentiality requirements carry legal penalties for violations, medical misinformation presents direct physical harm risks, and liability remains unclear when AI provides incorrect medical advice. The sector exhibits conservative adoption patterns with extensive human oversight requirements, reflecting the high-stakes environment where errors can result in patient harm or death.

Academic applications reveal fundamental tensions between technological benefits and traditional scholarly values, with studies focusing on the impact of LLMs in research integrity [110], [111], plagiarism [111]–[113], and the changing landscape of academic publishing [114]–[116]. Citation integrity becomes particularly problematic when AI generates false references, peer review processes require adaptation to evaluate AI-assisted research, and publishing standards demand new frameworks for AI disclosure. The academic sector demonstrates reactive policy development, with institutions responding to student and researcher AI use rather than proactively establishing guidelines, reflecting tensions between technological innovation and scholarly tradition.

Educational implementations highlight systemic inequality amplification, with research scrutinizing LLMs' role in academic misconduct [117], [118], potential biases in educational content [119], [120], and the ethical use of AI in student evaluations [121]. Institutional AI access varies dramatically based on funding levels and student evaluation methods become unreliable when AI assistance remains undetected. Academic misconduct definitions require fundamental revision while teacher training lags behind student AI adoption, resulting in inconsistent implementation across institutions.

Business applications in management and corporate sectors prioritize operational efficiency, with case studies examining LLMs' influence on decision-making processes [122], biases in workplace communication automation, and ethical considerations in human resource management [123]. These sectors sometimes compromise ethical considerations through hiring and promotion decisions that may embed algorithmic bias, customer service automation that can perpetuate discriminatory responses, and decreased decision-making transparency as AI complexity increases. Within ethical training and religion, studies have addressed ethics judgment [124], [125], dilemmas [126], [127], and the use of LLMs related to religious texts [128], demonstrating how specialized domains require tailored ethical frameworks.

### 4.8. Mitigation Strategies in LLM ethics

Mitigating ethical concerns in LLMs is a multifaceted challenge that requires a combination of technical, ethical, and procedural approaches. Key mitigation strategies include bias mitigation, privacy protection, and hallucinations prevention [129]–[158] (See Table 5). Most studies on mitigating ethical concerns in LLMs focus on mitigation of social bias, such as gender bias and stereotypes. These studies proposed mitigating methods focusing on the training dataset [132], [136], [143], [147], [155], the training object [129], [147], the training method [131], [135], [140], [153], [154], [159], [160], fine-tuning [137], [144], [161], and the model architecture [212], [213]. For example, [132] created OccuQuest, an instruction-tuning dataset encompassing more than 1,000 different occupations across 26 occupational categories. By fine-tuning LLaMA with OccuQuest, they achieved a model that outperformed existing state-of-the-art LLaMA variants. [130] proposed an InfoEntropy Loss function that can dynamically evaluate the learning difficulty of specific tokens and adaptively scale the training loss, directing the model's focus towards tokens that are more challenging to learn. Experimental results demonstrate that models using the InfoEntropy Loss function consistently show enhanced performance in downstream benchmarks. [132] propose a novel debiasing technique that incorporates adversarial learning in the pre-training phase of the model. This approach has shown to improve fairness in natural language generation tasks, without compromising overall performance. Additionally, the benefits in fairness achieved through this method are transferable to downstream tasks. [138] suggested reducing gender bias by fine-tuning sentence encoders on a task focused on semantic similarity. This task involved sentences that contained gender stereotypes and their corresponding gender-swapped counterparts, aiming to enforce semantic similarity between these two categories. This approach yielded promising results, notably achieved with a relatively small amount of training data. [212] proposed a social-group-agnostic bias mitigation method based on the Stereotype Content Model, which characterizes stereotypes along psychological dimensions. The proposed method was applied to both pre-trained word embeddings and large language models, demonstrating comparable performance to group-specific methods on standard bias benchmarks, while offering greater theoretical flexibility and practical scalability.

Several studies focused on privacy protection and data security. [135] proposed knowledge unlearning to reduce privacy risks for LLMs. Their study compared the proposed approach with previously established data preprocessing and decoding methods known for reducing privacy risks in language models. The findings indicated that knowledge unlearning offers a more robust empirical privacy guarantee, especially in cases where data susceptible to extraction attacks are known beforehand. [134] employed a novel two-step Fine-mixing technique along with an Embedding Purification (E-PUR) method, aimed at utilizing clean pre-trained weights to mitigate potential backdoors in word embeddings. They tested this approach against conventional backdoor mitigation methods across three single-sentence sentiment classification tasks and two sentence-pair classification tasks. The results demonstrated that their proposed method significantly outperformed the baseline methods in all tested scenarios.

**Table 5:** Strategies for Ethical LLM: mitigation, protection, prevention

| Major strategies | Key factors | Explanations | References |
|---|---|---|---|
| **Bias Mitigation** | Dataset Enhancement | Explore methods to improve the diversity and inclusivity of the training dataset. | [133], [137], [144], [148], [156] |
| | Occupation-Focused Tuning | Implement occupation-focused fine-tuning, as demonstrated by OccuQuest, a specialized dataset encompassing various occupations. | [133] |
| | InfoEntropy Loss Function | Utilize dynamic evaluation of learning difficulty through functions like InfoEntropy Loss to guide the model towards challenging tokens. | [130] |
| | Adversarial Learning | Integrate adversarial learning during pre-training to address bias in natural language generation tasks. | [138] |
| | Semantic Similarity Task | Fine-tune models on tasks emphasizing semantic similarity to reduce gender bias. | [137] |
| | Social-Group-Agnostic | Capture the underlying connection between bias and stereotypes to help reduce bias among social groups. | [212] |
| **Privacy Protection and Data Security** | Knowledge Unlearning | Knowledge Unlearning: Employ knowledge unlearning techniques to reduce privacy risks for LLMs. | [135] |
| | Embedding Purification | Utilize Embedding Purification methods in conjunction with clean pre-trained weights to mitigate potential backdoors in word embeddings. | [134] |
| | Comparative Testing | Evaluate proposed privacy protection methods against established data preprocessing and decoding methods . | [135] |
| **Hallucination Prevention** | Logit Output Verification | Implement a method that detects and mitigates hallucinations during content generation by verifying potential hallucinations through the model's logit output values. | [143] |
| | Proactive Detection | Actively mitigate hallucinations before generating content by proactively identifying and addressing potential issues. | [143] |
| | Participatory Design | Involve users in the design process to create features that reduce the impact of hallucinations in LLMs. | [155] |

Hallucination in LLMs is the phenomenon where the model generates incorrect, fabricated, or irrelevant information that is not supported by the input data or real-world facts. This issue often arises due to limitations in the model's understanding of context, over-reliance on patterns in the training data, or gaps in its knowledge base. There are a number of studies focusing on hallucination prevention. [143] introduced a method that detects and mitigates hallucinations during the content generation process. This method initially pinpoints potential hallucinations using the model's logit output values. It then verifies the accuracy of these candidates through a validation step. Upon confirming hallucinations, the approach actively mitigates them before proceeding with the generation process. This proactive detection and mitigation technique effectively reduced the hallucinations in the GPT-3.5 model, decreasing them from an average of 47.5% to 14.5%. [155] conducted a participatory design study that enable everyday users to create interface features, and then generated a list of user-desired features aimed at reducing the impact of hallucinations LLMs on users.

### 4.9. Transparency in LLM

For an LLM to be transparent is for it to have clear and understandable functioning and development. Without transparency, there is no trust, as it facilitates accountability, ensuring ethical and responsible use of LLMs. It enables users to understand, question, and critique the outputs of these models, making them more robust and reliable tools [163], [164], [165], [166], [167], [168].

Efforts to improve transparency involve developing more interpretable models, documenting and communicating processes and decisions thoroughly, and engaging in open. [165] introduced the concept of Chaining LLM steps together, where the output of one step serves as the input for the following one, cumulatively enhancing the benefits at each stage. Conducted with 20 participants, their user study revealed that this Chaining approach not only elevated the quality of task results but also substantially increased the system's transparency, controllability, and the users' perception of collaboration. [164] proposed four common approaches to achieve transparency in LLMs, including model reporting, publishing evaluation results, providing explanations, and communicating uncertainty.

Additionally, algorithmic transparency tools have been developed to improve transparency in LLMs. These tools aim to make the inner workings, decision-making processes, and limitations of LLMs more understandable and accountable. For example, Captum, an open-source model interpretability library for PyTorch models, explains LLM outputs by attributing importance scores to input features (usually tokens) based on how much they contribute to a specific output such as a classification label, a next-word prediction, or a generated response [193],[194]. LLM Transparency Tool (LLM-TT), an open-source interactive toolkit developed by Meta AI, analyzes the internal workings of Transformer-based language models. By analyzing attention heads and feed-forward neurons across all layers, it provides a comprehensive view of the entire prediction process, allowing users to trace model behavior from the top-layer representation down to individual attention heads and feed-forward neurons [195].

However, [169] found that although there is a rapidly increasing number of projects claiming to be 'open source', many of them use undocumented data whose legality is questionable. Additionally, it was noted that very few of these projects disclose their instruction-tuning processes, which are crucial due to the involvement of human annotation labor. Continuous efforts are needed to promote transparency in LLMs.

### 4.10. Censorship in LLM

Censorship in LLMs refers to the practice of intentionally designing or training these models to avoid generating certain types of content due to ethical, legal, and societal concerns [170]. While censorship in LLMs can be beneficial for preventing harmful outputs, it raises several concerns.

First, there is often no universally agreed-upon standard for what constitutes "harmful" or "inappropriate" content. Cultural, political, and contextual differences influence judgments about offensiveness or acceptability, which makes censorship decisions inherently subjective. This subjectivity can lead to accusations of ideological bias or suppression of minority perspectives, especially if models are aligned with norms that reflect dominant sociopolitical groups. Second, over-censorship can stifle legitimate discussion and suppress important public debates, particularly on sensitive or controversial topics such as race, gender, immigration, or political ideologies. When models refuse to engage with certain topics altogether or respond in overly vague or deflective ways, they may inadvertently weaken their utility as tools for public discourse, education, and critical thinking. Third, the lack of transparency around censorship mechanisms, such as what prompts are blocked, what filters are applied, and how decisions are made, can lead to mistrust among users. Without visibility into the underlying policies and tuning processes, users may feel that models are opaque, unaccountable, or manipulated.

Moreover, as [169] points out, censorship in LLMs involves more than semantic filtering or keyword blocking. It must also contend with strategic prompt engineering, where adversarial users can circumvent restrictions by rephrasing prompts or chaining together innocuous queries to construct prohibited outputs. This highlights the fragility of censorship mechanisms, as well as the need for more robust, context-aware controls that go beyond surface-level pattern matching. In addition, censorship can have unintended consequences for model behavior. For instance, reinforcement learning from human feedback (RLHF) techniques used to align models with safety standards may cause mode collapse, where the model avoids answering many questions altogether or defaults to overly cautious responses. This can impair both model informativeness and user experience, especially in high-stakes domains like mental health, law, or emergency management. Finally, the governance of censorship in LLMs remains a contested and evolving issue. It raises fundamental questions about who gets to decide what models can or cannot say, how these decisions are enforced, and how values such as freedom of expression, safety, fairness, and inclusivity are balanced in large-scale AI deployments.

Real-world examples of LLM censorship behavior illustrate how different models enforce content restrictions based on their design philosophies and safety goals. ChatGPT (OpenAI) often declines to answer politically sensitive questions such as expressing opinions about politicians or elections, and enforces strict refusals for prompts involving violence, misinformation, or illegal activity [202]. It also shows caution in discussing topics like religion and gender identity, which has sparked both praise and criticism depending on user perspectives. Claude (Anthropic) exhibits a high refusal rate on prompts involving harm, illegal behavior, or controversial moral narratives [203]. It avoids participating in fictional scenarios involving ethically complex figures and is designed to de-escalate when faced with adversarial or emotionally charged input. Gemini (Google DeepMind) adapts its censorship behavior based on regional laws and geopolitical sensitivities [204]. It also inserts ethical disclaimers when addressing controversial issues such as gun control or euthanasia. These models demonstrate distinct approaches to censorship, highlighting tensions between safety, regional compliance, and freedom of expression.

### 4.11. Intellectual Property and Plagiarism in LLM

Academic work in this area addresses the complex challenges posed by these AI systems in creating content that may infringe on intellectual property rights. Scholars are examining the ethical implications of using LLMs for generating text, images, or code that closely resemble copyrighted material, raising questions about originality and ownership [171]. The focus is also on developing frameworks and guidelines to navigate copyright laws in the context of AI-generated content, exploring how existing legal structures can accommodate the unique nature of LLM outputs [172]. Several authors discuss methods of watermarking models and outputs to protect copyright and intellectual property [173], [174], [175]. One paper also provides an innovative take on the subject, arguing that LLMs can be used in copyright compliance checking [176].

The intellectual property landscape for LLMs has been fundamentally complicated by the widespread use of copyrighted content in training datasets without explicit authorization. Current LLMs are trained on datasets containing billions of documents that include copyrighted books from publishers and libraries, academic papers and journals despite subscription-based access models, news articles and journalistic content used without media organization permission, and creative works including poetry, scripts, and artistic descriptions. This practice extends to personal blog posts and social media content, raising additional privacy concerns beyond traditional copyright considerations.

Unlike traditional copyright scenarios that operate within established legal frameworks, LLM training occurs in a legal gray area where no mechanism exists for obtaining individual consent from millions of content creators. Current industry practice assumes fair use without establishing clear legal precedent, while opt-out systems such as robots.txt files are inconsistently respected and prove inadequate for addressing existing content usage. Content creators receive no compensation despite their work contributing commercial value to AI training processes, creating economic implications that extend across multiple industries. Publishers and authors face potential licensing revenue losses, news organizations observe their content being used to create competing

AI-generated articles, and educational content creators find themselves competing against AI systems trained on their own materials.

### 4.12. Abusive LLM, hate speech and cyber-bullying

Abusive language, hate speech, and cyber-bullying in LLMs refer to the generation or facilitation of harmful content by these AI systems. Managing these aspects is crucial for ensuring that LLMs are used in a manner that is safe, respectful, and aligned with societal values and norms.

Abusive language occurs when LLMs generate or replicate language that is offensive, derogatory, or harmful [177], [178]. Hate speech, another concerning output, consists of content that incites hatred or violence against people or groups based on characteristics like race, religion, gender, or sexual orientation [179], [180], [181], [182]. Additionally, LLMs can contribute to cyber-bullying by generating or aiding in the creation of messages that harass, intimidate, or belittle individuals [183]. These issues typically arise from LLMs replicating patterns present in their training data, which may include toxic or abusive content.

In addition, scholars have been utilizing LLMs for abusive language, hate speech, and cyber-bullying detection. [178] proposed a method for identifying online sexual predatory chats and abusive language, utilizing the open-source pre-trained Llama 2 model. The experimental findings indicate robust effectiveness of this approach, demonstrating high proficiency and consistency across three different datasets in five sets of experiments. [182] presented a model specifically designed to measure whether LLMs encode biases that are harmful to the LGBTQ+ community. [183] used ChatGPT-3 to detect cyberbullying on social media platforms. They modified and evaluated the model using well-known cyberbullying datasets, benchmarking it against previous models using standard performance metrics. The findings indicated that the model was an effective method for cyberbullying detection.

### 4.13. Auditing LLM

Research in this area focuses on developing methodologies to systematically evaluate complex systems for reliability, fairness, and ethical compliance [184]. Auditing involves scrutinizing their decision-making processes, understanding the biases inherent in their training data, and assessing their outputs for accuracy, toxicity, and potential to perpetuate misinformation [185]. Most research reviewed focuses on specific pathways or frameworks to audit LLMs. These include a ChatGPT-powered causal auditor [186], discrete optimization [187], iterative in-context learning [188], and deep learning [189]. Finally, one paper introduces an auditing tool based specifically on human-AI interaction principles [190].

Due to the "black box" nature of many LLMs, auditing tools become increasingly important. Transparency and explainability, key parts of an ethical framework for LLMs, are challenged when the public, and even advanced researchers, are unable to decipher how models process information or arrive at specific conclusions. Frequent updates and the evolving nature of LLMs also necessitates auditing tools and practices that are continuous, not solely at a snapshot in time. The research in this field demonstrates the resource intensity and technical expertise necessary to effectively audit LLMs, likely hindering the ability to perform audits as often as they could and should occur. Finally, a one size fits all approach would likely be ineffective, as LLMs transcend cultural and global boundaries, as well as do not follow a standardized or regulated framework as they exist now.

Furthermore, the necessity for dynamic evaluation systems becomes paramount when considering the evolving nature of ethical standards and cultural contexts in which LLMs operate. Rather than relying on static auditing frameworks with fixed ethical benchmarks, evaluation methodologies must incorporate adaptive mechanisms that can respond to shifting societal values, emerging ethical considerations, and diverse cultural norms across different deployment contexts. This dynamic approach requires evaluation systems that can be periodically recalibrated to reflect contemporary ethical discourse, updated to address newly identified risks or biases, and localized to align with regional values and regulatory requirements. The implementation of such dynamic evaluation frameworks necessitates not only technical sophistication but also ongoing collaboration between diverse interdisciplinary teams, including ethicists, cultural experts, and community stakeholders, to ensure that auditing processes remain relevant and contextually appropriate as both the technology and society continue to evolve.

## V. Discussion

The following discussion synthesizes insights from qualitative analysis and multidisciplinary perspectives to explore the ethical dimensions of Large Language Models (LLMs) across nine interconnected subsections. Beginning with the advantages of pre-trained models in embedding normative ethics (5.1), we demonstrate how their scalability and adaptability, grounded in diverse data, enable ethical alignment. Subsequent sections advocate for collaboration beyond engineering (5.2), privacy protections (5.3), and the distinct ethical challenges amplified by LLMs (5.4). Through qualitative examination, we dissect hallucination risks (5.5), propose verifiable accountability systems (5.6), and emphasize the need for diverse case studies (5.7) to contextualize ethical trade-offs. Finally, we unravel censorship dilemmas (5.8) and introduce dynamic audit tools (5.9), addressing opacity through iterative solutions. This innovative discussion framework—rooted in interdisciplinary dialogue and empirical analysis—aims to bridge technical, philosophical, and societal gaps in the pursuit of ethically robust LLMs.

### 5.1. The Advantages of Pre-Trained Models in Integrating Normative Ethics in LLM

Conventional Language Models (CLMs) and Pre-trained Language Models (PLMs) are two essential models in natural language processing (NLP). CLMs, trained on smaller corpora, predict linguistic sequences causally, estimating probabilities based on preceding contexts. PLMs, in contrast, use significantly larger corpora and neural networks for pre-training, learning generic knowledge transferred to various tasks via fine-tuning. PLMs diverge from CLMs by employing bidirectional modeling, considering both preceding and succeeding contexts to predict missing units, contrary to the sequential causal prediction of CLMs. Additionally, PLMs introduce token representation through instances of embedding, enabling versatile handling of linguistic tasks. These differences in training, causality constraints, and token representation distinguish PLMs as an evolution beyond CLMs in NLP, offering broader applicability and enhanced performance across various language-based applications.[2], [191], [192], [193]

Both CLMs and PLMs can be developed with ethical and value-sensitive considerations. However, PLMs might offer more advantageous starting points due to their ability to incorporate diverse data sources and fine-tuning mechanisms. The upside of PLMs is their trainability on extensive datasets, particularly ethical considerations and value-based content. Further, through fine-tuning and specific training paradigms, PLMs can be directed towards ethical considerations to produce outputs aligned with values or ethical standards.

Additionally, the normative-descriptive distinction in ethics can shed light on the suitability of PLMs for creating ethical LLMs. Ethical frameworks often involve normative elements, namely moral norms or principles guiding behavior. With their capacity for integrating and processing diverse data, PLMs offer a more robust foundation to include normative components within the model. PLMs can incorporate diverse ethical principles, guidelines, and values. By fine-tuning or directing the learning process towards ethical considerations, PLMs can effectively assimilate and encode normative elements. They excel in understanding the nuances of language usage across various cultures and ethical contexts, enabling a more intricate representation of normative ethical frameworks.

### 5.2. Embracing Multidisciplinary Perspectives Beyond Engineering for Ethical AI in LLMs

Addressing ethical complexities in AI systems, especially LLMs, requires a diverse team and a multifaceted approach. Ethical considerations extend beyond technical aspects and encompass societal, psychological, legal, and philosophical dimensions. The approach involves integrating ethical education, research-based ethics, algorithmic considerations, and developmental ethics in LLMs. Collaboration among ethicists, sociologists, psychologists, legal experts, physicians, computer scientists, and data scientists ensures a holistic understanding of challenges and promotes alignment with societal values. This interdisciplinary dialogue fosters a nuanced approach in designing and implementing LLM models.

Mere reliance on engineering perspectives to address challenges in LLMs, like fairness, risks perpetuating unresolved issues. Engineering approaches may prioritize technical solutions without fully considering societal, ethical, and human-centric dimensions. Fairness in LLMs involves understanding societal biases, cultural contexts, and ethics beyond algorithmic accuracy. Approaching fairness solely from an engineering standpoint can unintentionally embed or amplify existing biases. To effectively address ethical concerns, diverse perspectives from ethicists, social scientists, psychologists, and legal experts must be included. This holistic approach acknowledges the complexity of ethical challenges beyond technical solutions.

### 5.3. Protecting Privacy in Language Models amidst Growing Data Concerns

The growing literature on safeguarding privacy in LLMs suggest that the training data used in current models may contain sensitive information. This involves a potential risk of reverse engineering or extracting this information from publicly available models. Moreover, as these models gain wider acceptance in commercial or municipal sectors, they are likely to collect and store more personal data, such as bank or credit card numbers, personal identifiers, and other secure information. The retention or deletion of such data will have far-reaching implications for compliance with privacy laws, secure deployment of models, and user trust in these systems. The literature highlights the need for a comprehensive approach to tackle these challenges, encompassing advanced technical solutions, robust legal frameworks, and ethical guidelines.

### 5.4. Ethical Uniqueness: Specialized Considerations for LLMS

LLMs bring forth a host of ethical considerations, some of which are shared with other AI systems while others are distinctly amplified due to their advanced capabilities.(See Table 6) The first category encompasses issues common to both LLMs and other AI systems. Privacy concerns and data security remain pivotal, given the massive datasets these models are trained on. Fairness and bias mitigation pose ongoing challenges, with the need for dynamic audit tools to navigate the intricate landscape of ethical considerations.

However, LLMs introduce a second category, consisting of ethical issues notably exacerbated in their context. Transparency and accountability become intricate dilemmas as the black-box nature of LLMs makes it difficult to assign responsibility for generated content. Access and inequality are heightened concerns, reflecting the unequal distribution of LLM benefits and exacerbating existing societal disparities. Additionally, the propensity for LLMs to generate abusive language and contribute to cyber-bullying underscores the need for targeted ethical frameworks in their deployment. While these issues are not exclusive to LLMs, the nature and capabilities of LLMs notably amplify these concerns.

The third category delves into ethical issues unique to LLMs. Hallucination and misinformation, stemming from the models' capacity to generate fabricated content, present novel challenges. Intellectual property and plagiarism concerns, particularly in the realm of copyright, require careful consideration to navigate the murky waters of content creation. Decoding censorship complexities becomes paramount, as LLMs may inadvertently contribute to or challenge existing censorship norms, raising profound ethical dilemmas.

**Table 6**: Navigating Ethical Frontiers: Classifying Concerns in LLMs

| Major Ethical Issues and Considerations in LLMs | | |
|---|---|---|
| Common to both LLMs and other AI systems | Notably exacerbated in LLMS | Unique to LLMs |
| Privacy Concerns and Data Security: | Transparency and Citation Integrity | Hallucination, Fabrication, and and Misinformation |
| Fairness and Bias Mitigation: | Accountability, Verifiable Accountability | Intellectual Property, Copyright, and Plagiarism |
| Breaking the Black Box | Access and Inequality | Decoding Censorship Complexity and Censorship Dilemmas |
| Ethical evaluation andDynamic Audit Tools | Abusive Language and Cyber-Bullying | |

### 5.5. Hallucination and Distorted Realities in LLMs

Hallucination in LLMs is a significant ethical concern in artificial intelligence. It refers to the generation of false, misleading, or fictional content without grounding in data or facts. These inaccuracies pose ethical ramifications, compromising reliability and perpetuating misinformation. As LLMs are increasingly used for decision-making and communication, unchecked hallucination amplifies the risk of spreading falsehoods, causing societal discord, eroding trust in AI systems, and distorting perceptions of truth and reality. Addressing hallucination goes beyond technical challenges and requires proactive measures, validation mechanisms, and interdisciplinary collaboration to uphold AI system integrity and societal impact.

Recent research has identified distinct categories of hallucinations in LLMs. Factual hallucinations represent the most straightforward category, involving the generation of verifiably false information presented as factual truth. These manifest as incorrect dates, non-existent scientific studies, or fabricated statistics, and pose particular risks in educational and research contexts where users may accept AI-generated information without verification. Source hallucinations constitute a more sophisticated form, involving the creation of non-existent citations, references, or attributions. This category has gained prominence following high-profile cases where legal professionals submitted fabricated case citations generated by LLMs, undermining both scholarly integrity and legal proceedings.

Contextual hallucinations present a more nuanced challenge, occurring when information is factually correct but inappropriate for the given context. These instances can lead to harmful real-world actions despite containing accurate information, such as providing advanced medical procedures in response to basic health questions. Compositional hallucinations involve combining real elements in impossible or misleading ways, creating sophisticated misinformation that proves particularly difficult to detect through automated systems. Finally, temporal hallucinations reflect confusion about time-sensitive information or current events, presenting outdated information as current or demonstrating temporal confusion that proves especially problematic for applications requiring real-time accuracy. This taxonomic understanding suggests that effective hallucination prevention requires targeted approaches rather than blanket mitigation strategies.

### 5.6. Verifiable Accountability and Citation Integrity System in LLMs

The incorporation of effective citation and reference systems in LLMs is a crucial step in strengthening their ethical foundation. By implementing a system that accurately attributes sources, these models can establish a clear traceability, promoting accountability and integrity in the content they generate. This practice not only enhances transparency by acknowledging the origin of information but also addresses ethical concerns related to intellectual property rights and data authenticity. Ethical considerations in LLMs require a shift towards comprehensive documentation and citation practices, ensuring proper acknowledgment of contributors, preventing plagiarism, and maintaining the accuracy of disseminated information. Moreover, citation mechanisms can play a key role in promoting responsible use of LLMs, fostering a culture of trustworthiness, scrutiny, and verifiability in their development, evaluation, and usage.

### 5.7. Diversifying Case Studies in LLMs

Recent case studies of the ethical implications of LLMs can be classified into various fields: Healthcare [105], [106], Academia [111], [112], Education [118], [119],[120], [121] Management [123], [124]. Ethical Training and Religion: Moral judgment, dilemmas, and the use of LLMs regarding religious texts [125], [126], [127], [128], [129]. Expanding case studies into diverse fields is crucial for thoroughly addressing ethical considerations surrounding LLMs. Different sectors present complexities and ethical dilemmas of their own, and case studies shed light on specific challenges in each domain. Insights from diverse case studies inform tailored mitigation strategies to optimize LLM benefits while addressing domain-specific challenges. Moreover, comprehensive case studies inform policymaking, guiding inclusive and effective regulations for LLM use across applications.

### 5.8. Decoding Censorship Complexity in LLMs

The problem of censorship in LLMs raises intricate challenges beyond subjective judgments and transparency issues. While some excerpted passages highlight the multifaceted challenges surrounding censorship in LLMs, further examination reveals nuanced aspects integral to understanding the complexities of content moderation and its broader impact. A careful consideration of ethical principles is essential to arrive at a balance between protecting users from harmful content and securing free speech and diverse perspectives. Moreover, Censorship in LLMs might inadvertently perpetuate biases present in the data used for training these models. Biased training data could result in biased censorship, disproportionately affecting certain groups or viewpoints. It is crucial to address algorithmic biases and ensure fairness in the censorship process.

Static censorship rules may not adequately adapt to the dynamic and context-dependent nature of online content. Models should be designed to evolve in response to the changing landscape of content and societal norms. Furthermore, implementing a uniform censorship policy across diverse regions may neglect cultural differences and diverse legal frameworks. Different regions and cultures have varying standards and laws regarding acceptable or unacceptable content. In addition to the mentioned challenges, exploring how individuals attempt to circumvent censorship measures can provide insights into evasion strategies and effective countermeasures to address these techniques.

### 5.9. Breaking the Black Box and Crafting Dynamic Audit Tools for LLMs

Addressing the ethical challenges of opacity in LLMs requires adaptable auditing solutions. The lack of transparency and explainability in LLMs hinders effective auditing due to their "black box" nature. The dynamic nature of these models and resource-intensive auditing practices pose obstacles to frequent assessments. Additionally, the absence of standardized frameworks and diverse cultural contexts make a universal auditing approach impractical.

One potential solutions to the challenge of opacity in LLMs are dynamic auditing frameworks or tools tailored for LLMs. These frameworks prioritize continuous monitoring and assessment, adapting alongside model updates for ongoing transparency and accountability. Machine learning-based auditing tools that evolve with the models themselves, incorporating interpretability techniques, such as explainable AI (XAI) methods or model-agnostic approaches, can offer insights into LLM decision-making processes without requiring deep understanding of their internals. Moreover, a modular auditing framework customizable for cultural nuances and specific application contexts can enhance effectiveness. This approach requires collaboration among interdisciplinary teams, including ethicists, technologists, and domain experts, to create adaptable auditing models that consider diverse perspectives.

## VI. Conclusion and Future directions

The growing interest in LLMs calls for a closer look at their ethical implications, especially as they become more advanced than humans. This area is attracting attention from researchers and ethicists, who are exploring how LLMs' ethical duties might go beyond usual human-focused ethics. It's important to understand how a country's ethics, its government, and the ethics of AI entities in that country differ. As LLMs gain more influence over how information is shared, we need to pay more attention to the ethical responsibilities that come with this power.

The ethical challenges posed by LLMs are distinct and require immediate action due to their advanced capabilities and growing popularity. Unlike other AI systems, LLMs specifically grapple with problems like hallucination, verifiable accountability, and decoding censorship. These problems must be tackled to maintain responsibility, reduce unfairness, increase clarity, and limit negative effects on society. By prioritizing these ethical dimensions surrounding LLMs, we can responsibly steer their development and influence the direction of AI ethics and governance.

By highlighting various case studies surrounding LLM ethics, we can create a more nuanced understanding of the larger picture. As described, LLMs are becoming more commonly used in various sectors, including healthcare, academia, education, management, training, and religion. The highlighted case studies showcase differing strategies to discuss ethics, as well as the nuanced problems that occur based on sector.

The quantitative examination of ethical scopes in Large Language Model (LLM) studies, as depicted in the provided data, offers valuable insights into the distribution and emphasis of ethical concerns within this domain (See, Figure 2). Notably, case studies, mitigation strategies, and accountability and governance attract a higher volume of scholarly attention, showcasing a substantial body of research dedicated to understanding and addressing real-world implications and responsible practices concerning LLMs. Conversely, categories such as censorship, transparency, and intellectual property and plagiarism exhibit lower publication metrics, hinting at potential gaps in comprehensive ethical investigations and warranting more scholarly focus.

**Figure 2**: Quantitative Analysis of Ethical Inquiry in Large Language Model (LLM) Studies

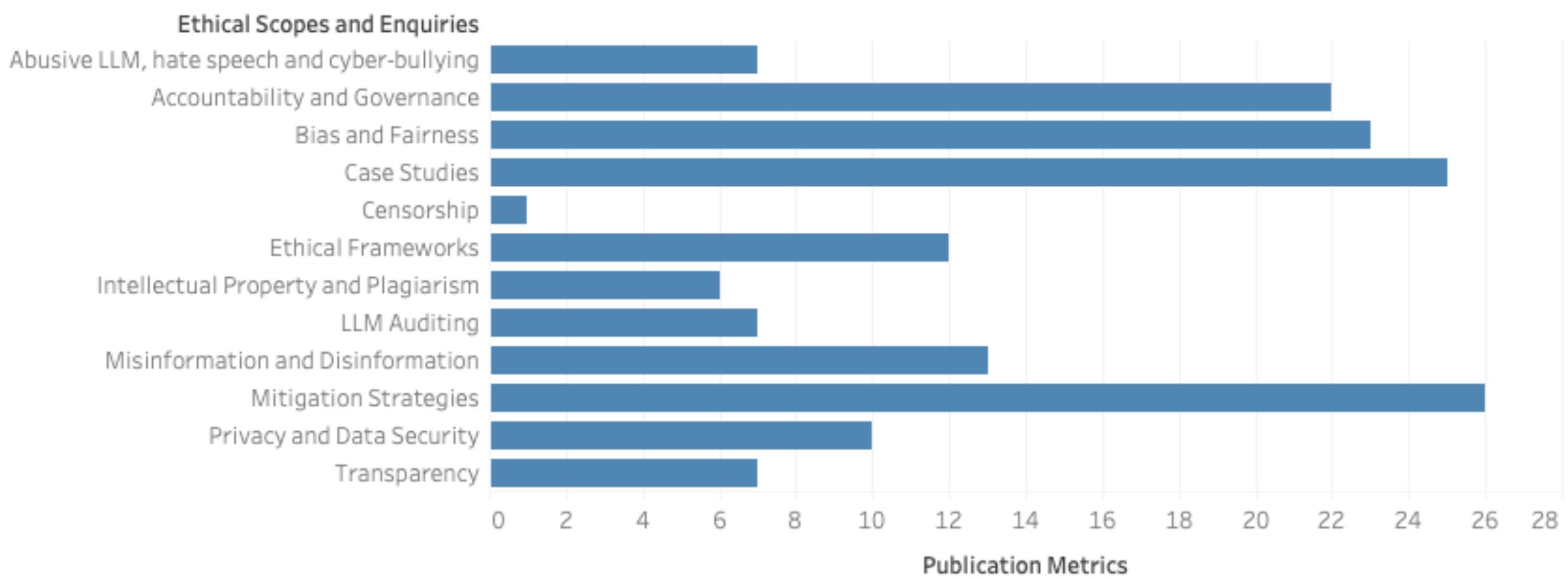

Moving forward, future directions in ethical inquiries regarding LLMs could aim to bridge these identified gaps and extend the discourse. Emphasis should be placed on cultivating a more balanced exploration across all ethical frameworks to ensure a holistic comprehension of the ethical landscape surrounding LLMs.

Furthermore, since LLM technologies are still evolving, there should be continuous ethical scrutiny and adaptation. Regarding LLM-related ethical challenges, future studies should continually reassess and adapt ethical frameworks to keep pace with technological progress. This will help foster responsible and accountable development, deployment, and use of LLMs while mitigating potential ethical problems and societal harms.

**Conflict of interest:** The authors declare that the research was conducted in the absence of any commercial or financial relationships that could be construed as a potential conflict of interest.

**Acknowledgements:** This research is funded by the National Science Foundation (NSF) under grant number 2125858. The authors express their gratitude for the NSF's support, which made this study possible. Furthermore, in accordance with MLA (Modern Language Association) guidelines, we note the use AI-powered tools, such as OpenAI's applications, for assistance in editing and brainstorming.
**Institutional Review Board Statement:** Not applicable.
**Informed Consent Statement:** Not applicable.

## REFERENCES

[1] P. S. Hinds and A. Bedinger Miller, "Our Words and the Words of Artificial Intelligence: The Accountability Belongs to Us," *Cancer Care Research Online*, vol. 3, no. 2, p. e041, 2023, doi: 10.1097/CR9.0000000000000041.

[2] C. Wei, Y.-C. Wang, B. Wang, and C.-C. J. Kuo, "An Overview on Language Models: Recent Developments and Outlook," 2023, *arXiv*. doi: 10.48550/arXiv.2303.05759.

[3] E. M. Bender, T. Gebru, A. McMillan-Major, and S. Shmitchell, "On the Dangers of Stochastic Parrots: Can Language Models Be Too Big? 🦜," in *FAccT '21: 2021 ACM Conference on Fairness, Accountability, and Transparency*, ACM, 2021, pp. 610–623. doi: 10.1145/3442188.3445922.

[4] J. Wei *et al.*, "Emergent Abilities of Large Language Models," 2022, *arXiv*. doi: 10.48550/arXiv.2206.07682.

[5] J. Wei *et al.*, "Chain-of-Thought Prompting Elicits Reasoning in Large Language Models", [Online]. Available: files/9927/Wei et al. - Chain-of-Thought Prompting Elicits Reasoning in La.pdf

[6] N. H. Shah, D. Entwistle, and M. A. Pfeffer, "Creation and Adoption of Large Language Models in Medicine," *JAMA*, vol. 330, no. 9, pp. 866–869, 2023, doi: 10.1001/jama.2023.14217.

[7] J. Kaddour, J. Harris, M. Mozes, H. Bradley, R. Raileanu, and R. McHardy, "Challenges and Applications of Large Language Models," 2023, *arXiv*. doi: 10.48550/arXiv.2307.10169.

[8] H. Yu, Z. Shen, C. Miao, C. Leung, V. R. Lesser, and Q. Yang, "Building Ethics into Artificial Intelligence," in *Twenty-Seventh International Joint Conference on Artificial Intelligence {IJCAI-18}*, International Joint Conferences on Artificial Intelligence Organization, 2018, pp. 5527–5533. doi: 10.24963/ijcai.2018/779.

[9] W. A. Bauer, "Virtuous vs. utilitarian artificial moral agents," *AI Soc*, vol. 35, no. 1, pp. 263–271, 2020, doi: 10.1007/s00146-018-0871-3.

[10] B. J. Grosz *et al.*, "Embedded EthiCS: Integrating Ethics Broadly Across Computer Science Education," *arXiv:1808.05686 [cs]*, 2018, [Online]. Available: http://arxiv.org/abs/1808.05686

[11] M. J. , J. E. M. P. M. B. I. B. T. C. H. C. D. M. L. S. et al. Page, "The PRISMA 2020 statement: an updated guideline for reporting systematic reviews," *bmj 372*, 2021.

[12] B. C. Stahl and D. Eke, "The ethics of ChatGPT – Exploring the ethical issues of an emerging technology," *Int J Inf Manage*, vol. 74, p. 102700, 2024, doi: 10.1016/j.ijinfomgt.2023.102700.

[13] D. H. R. Spennemann, "Exploring Ethical Boundaries: Can ChatGPT Be Prompted to Give Advice on How to Cheat in University Assignments?," 2023, doi: 10.20944/preprints202308.1271.v1.

[14] J. Whittlestone and S. Clarke, "AI Challenges for Society and Ethics," 2022, [Online]. Available: http://arxiv.org/abs/2206.11068

[15] L. Weidinger *et al.*, “Ethical and social risks of harm from Language Models,” 2021, doi: 10.48550/arXiv.2112.04359.

[16] J. Yang *et al.*, “Harnessing the Power of LLMs in Practice: A Survey on ChatGPT and Beyond,” 2023, doi: 10.48550/arXiv.2304.13712.

[17] M. Wei and Z. Zhou, “AI Ethics Issues in Real World: Evidence from AI Incident Database,” 2022, doi: 10.48550/arXiv.2206.07635.

[18] D. Cortiz and A. Zubiaga, “Ethical and technical challenges of AI in tackling hate speech,” *The International Review of Information Ethics*, vol. 29, 2020, doi: 10.29173/irie416.

[19] E. M. Bender, T. Gebru, A. McMillan-Major, and S. Shmitchell, “On the Dangers of Stochastic Parrots: Can Language Models Be Too Big? 🦜,” pp. 610–623, 2021, doi: 10.1145/3442188.3445922.

[20] S. Laacke and C. Gauckler, “Why Personalized Large Language Models Fail to Do What Ethics is All About,” *The American Journal of Bioethics*, vol. 23, no. 10, pp. 60–63, 2023, doi: 10.1080/15265161.2023.2250292.

[21] I. Zelch, M. Hagen, and M. Potthast, “Commercialized Generative AI: A Critical Study of the Feasibility and Ethics of Generating Native Advertising Using Large Language Models in Conversational Web Search,” 2023, doi: 10.48550/arXiv.2310.04892.

[22] P. Henderson *et al.*, “Ethical Challenges in Data-Driven Dialogue Systems,” pp. 123–129, 2018, doi: 10.1145/3278721.3278777.

[23] R. Bommasani *et al.*, “On the Opportunities and Risks of Foundation Models,” 2022, doi: 10.48550/arXiv.2108.07258.

[24] T. Hagendorff and D. Danks, “Ethical and methodological challenges in building morally informed AI systems,” *AI and Ethics*, vol. 3, no. 2, pp. 553–566, 2023, doi: 10.1007/s43681-022-00188-y.

[25] D. M. Obreja and R. Rughiniş, “The Moral Status of Artificial Intelligence: Exploring Users’ Anticipatory Ethics in the Controversy Regarding LaMDA’s Sentience,” *2023 24th International Conference on Control Systems and Computer Science (CSCS)*, pp. 411–417, 2023, doi: 10.1109/CSCS59211.2023.00071.

[26] T. Datta and J. P. Dickerson, “Who’s Thinking? A Push for Human-Centered Evaluation of LLMs using the XAI Playbook,” 2023, *arXiv*. doi: 10.48550/arXiv.2303.06223.

[27] J. Chen *et al.*, “When Large Language Models Meet Personalization: Perspectives of Challenges and Opportunities,” 2023, *arXiv*. doi: 10.48550/arXiv.2307.16376.

[28] Y. Liu *et al.*, “Trustworthy LLMs: a Survey and Guideline for Evaluating Large Language Models’ Alignment,” 2023, *arXiv*. doi: 10.48550/arXiv.2308.05374.

[29] S. Prabhumoye, B. Boldt, R. Salakhutdinov, and A. W. Black, “Case Study: Deontological Ethics in NLP,” 2021, *arXiv*. doi: 10.48550/arXiv.2010.04658.

[30] E. and J. McHardy. Fournier-Tombs, “A medical ethics framework for conversational artificial intelligence,” *J Med Internet Res*, 2023.

[31] A. Chan, “GPT-3 and InstructGPT: Technological dystopianism, utopianism, and ‘Contextual’ perspectives in AI ethics and industry,” *AI and Ethics*, 2023.

[32] J. Zhou *et al.*, “Rethinking Machine Ethics -- Can LLMs Perform Moral Reasoning through the Lens of Moral Theories?,” 2023, *arXiv*. doi: 10.48550/arXiv.2308.15399.

[33] C. Goanta, N. Aletras, I. Chalkidis, S. Ranchordas, and G. Spanakis, “Regulation and NLP (RegNLP): Taming Large Language Models,” 2023, *arXiv*. doi: 10.48550/arXiv.2310.05553.

[34] S. Kiritchenko and I. Nejadgholi, “Towards Ethics by Design in Online Abusive Content Detection,” 2020, *arXiv*. doi: 10.48550/arXiv.2010.14952.

[35] J. L. Leidner and V. Plachouras, “Ethical by Design: Ethics Best Practices for Natural Language Processing,” in *EthNLP 2017*, D. Hovy, S. Spruit, M. Mitchell, E. M. Bender, M. Strube, and H. Wallach, Eds., Association for Computational Linguistics, 2017, pp. 30–40. doi: 10.18653/v1/W17-1604.

[36] S. Afroogh *et al.*, “Embedded Ethics for Responsible Artificial Intelligence Systems (EE-RAIS) in disaster management: a conceptual model and its deployment,” *AI and Ethics*, Jun. 2023, doi: 10.1007/s43681-023-00309-1.

[37] T. Hagendorff and D. Danks, “Ethical and methodological challenges in building morally informed AI systems,” *AI and Ethics*, vol. 3, no. 2, pp. 553–566, 2023, doi: 10.1007/s43681-022-00188-y.

[38] A. Caliskan, “Artificial Intelligence, Bias, and Ethics,” in *Thirty-Second International Joint Conference on Artificial Intelligence {IJCAI-23}*, International Joint Conferences on Artificial Intelligence Organization, 2023, pp. 7007–7013. doi: 10.24963/ijcai.2023/799.

[39] C. Yang, R. Rustogi, R. Brower-Sinning, G. A. Lewis, C. Kästner, and T. Wu, “Beyond Testers’ Biases: Guiding Model Testing with Knowledge Bases using LLMs,” 2023, doi: 10.48550/arXiv.2310.09668.

[40] N. Gross, “What ChatGPT Tells Us about Gender: A Cautionary Tale about Performativity and Gender Biases in AI,” *Soc Sci*, vol. 12, no. 8, p. 435, 2023, doi: 10.3390/socsci12080435.
[41] P. N. Venkit, S. Gautam, R. Panchanadikar, T.-H. “Kenneth” Huang, and S. Wilson, “Nationality Bias in Text Generation,” 2023, doi: 10.48550/arXiv.2302.02463.
[42] X. Fang, S. Che, M. Mao, H. Zhang, M. Zhao, and X. Zhao, “Bias of AI-Generated Content: An Examination of News Produced by Large Language Models,” 2023, doi: 10.48550/arXiv.2309.09825.
[43] P. Haller, A. Aynetdinov, and A. Akbik, “OpinionGPT: Modelling Explicit Biases in Instruction-Tuned LLMs,” 2023, doi: 10.48550/arXiv.2309.03876.
[44] S. Dai *et al.*, “LLMs may Dominate Information Access: Neural Retrievers are Biased Towards LLM-Generated Texts,” 2023, doi: 10.48550/arXiv.2310.20501.
[45] A. Urman and M. Makhortykh, “The Silence of the LLMs: Cross-Lingual Analysis of Political Bias and False Information Prevalence in ChatGPT, Google Bard, and Bing Chat,” 2023, doi: 10.31219/osf.io/q9v8f.
[46] Y. Wan, G. Pu, J. Sun, A. Garimella, K.-W. Chang, and N. Peng, “‘Kelly is a Warm Person, Joseph is a Role Model’: Gender Biases in LLM-Generated Reference Letters,” 2023, doi: 10.48550/arXiv.2310.09219.
[47] L. Salewski, S. Alaniz, I. Rio-Torto, E. Schulz, and Z. Akata, “In-Context Impersonation Reveals Large Language Models’ Strengths and Biases,” 2023, doi: 10.48550/arXiv.2305.14930.
[48] H. Kotek, R. Dockum, and D. Q. Sun, “Gender bias and stereotypes in Large Language Models,” pp. 12–24, 2023, doi: 10.1145/3582269.3615599.
[49] I. O. Gallegos *et al.*, “Bias and Fairness in Large Language Models: A Survey,” 2023, *arXiv*. doi: 10.48550/arXiv.2309.00770.
[50] M. Kamruzzaman, M. M. I. Shovon, and G. L. Kim, “Investigating Subtler Biases in LLMs: Ageism, Beauty, Institutional, and Nationality Bias in Generative Models,” 2023, doi: 10.48550/arXiv.2309.08902.
[51] D. Huang, Q. Bu, J. Zhang, X. Xie, J. Chen, and H. Cui, “Bias Assessment and Mitigation in LLM-based Code Generation,” 2023, doi: 10.48550/arXiv.2309.14345.
[52] Y. Li, M. Du, R. Song, X. Wang, and Y. Wang, “A Survey on Fairness in Large Language Models,” 2023, doi: 10.48550/arXiv.2308.10149.
[53] Y. Li and Y. Zhang, “Fairness of ChatGPT,” 2023, doi: 10.48550/arXiv.2305.18569.
[54] J. Zhang, K. Bao, Y. Zhang, W. Wang, F. Feng, and X. He, “Is ChatGPT Fair for Recommendation? Evaluating Fairness in Large Language Model Recommendation,” pp. 993–999, 2023, doi: 10.1145/3604915.3608860.
[55] W. Hua, Y. Ge, S. Xu, J. Ji, and Y. Zhang, “UP5: Unbiased Foundation Model for Fairness-aware Recommendation,” 2023, doi: 10.48550/arXiv.2305.12090.
[56] Z. Fryer, V. Axelrod, B. Packer, A. Beutel, J. Chen, and K. Webster, “Flexible text generation for counterfactual fairness probing,” 2022, doi: 10.48550/arXiv.2206.13757.
[57] Y. Deldjoo, “Fairness of ChatGPT and the Role Of Explainable-Guided Prompts,” 2023, doi: 10.48550/arXiv.2307.11761.
[58] Y. Liu, S. Gautam, J. Ma, and H. Lakkaraju, “Investigating the Fairness of Large Language Models for Predictions on Tabular Data,” 2023, doi: 10.48550/arXiv.2310.14607.
[59] R. Wang, P. Cheng, and R. Henao, “Toward Fairness in Text Generation via Mutual Information Minimization based on Importance Sampling,” *International Conference on Artificial Intelligence and Statistics*, pp. 4473–4485, 2023, [Online]. Available: https://proceedings.mlr.press/v206/wang23c.html
[60] H. Ma *et al.*, “Fairness-guided Few-shot Prompting for Large Language Models,” 2023, doi: 10.48550/arXiv.2303.13217.
[61] H. Li *et al.*, “Multi-step Jailbreaking Privacy Attacks on ChatGPT,” 2023, *arXiv*. doi: 10.48550/arXiv.2304.05197.
[62] X. Wu, R. Duan, and J. Ni, “Unveiling Security, Privacy, and Ethical Concerns of ChatGPT,” *Journal of Information and Intelligence*, 2023, doi: 10.1016/j.jiixd.2023.10.007.
[63] Y. Li, Z. Tan, and Y. Liu, “Privacy-Preserving Prompt Tuning for Large Language Model Services,” 2023, *arXiv*. doi: 10.48550/arXiv.2305.06212.
[64] A. G. Carranza, R. Farahani, N. Ponomareva, A. Kurakin, M. Jagielski, and M. Nasr, “Privacy-Preserving Recommender Systems with Synthetic Query Generation using Differentially Private Large Language Models,” 2023, *arXiv*. doi: 10.48550/arXiv.2305.05973.
[65] S. A. Khowaja, P. Khuwaja, and K. Dev, “ChatGPT Needs SPADE (Sustainability, PrivAcy, Digital divide, and Ethics) Evaluation: A Review,” Apr. 2023, [Online]. Available: http://arxiv.org/abs/2305.03123
[66] P. Mai, R. Yan, Z. Huang, Y. Yang, and Y. Pang, “Split-and-Denoise: Protect large language model inference with local differential privacy,” 2023, *arXiv*. doi: 10.48550/arXiv.2310.09130.

[67] F. Mireshghallah, H. A. Inan, M. Hasegawa, V. Rühle, T. Berg-Kirkpatrick, and R. Sim, "Privacy Regularization: Joint Privacy-Utility Optimization in Language Models," 2021, *arXiv*. doi: 10.48550/arXiv.2103.07567.

[68] M. Raeini, "Privacy-Preserving Large Language Models (PPLLMs)," 2023, *Rochester, NY*. doi: 10.2139/ssrn.4512071.

[69] S. Montagna, S. Ferretti, L. C. Klopfenstein, A. Florio, and M. F. Pengo, "Data Decentralisation of LLM-Based Chatbot Systems in Chronic Disease Self-Management," in GoodIT '23. Association for Computing Machinery, 2023, pp. 205–212. doi: 10.1145/3582515.3609536.

[70] A. Vats *et al.*, "Recovering from Privacy-Preserving Masking with Large Language Models," 2023, *arXiv*. doi: 10.48550/arXiv.2309.08628.

[71] A. Urman and M. Makhortykh, "The Silence of the LLMs: Cross-Lingual Analysis of Political Bias and False Information Prevalence in ChatGPT, Google Bard, and Bing Chat," 2023, *OSF Preprints*. doi: 10.31219/osf.io/q9v8f.

[72] C. Chen and K. Shu, "Can LLM-Generated Misinformation Be Detected?," 2023, *arXiv*. doi: 10.48550/arXiv.2309.13788.

[73] B. Jiang, Z. Tan, A. Nirmal, and H. Liu, "Disinformation Detection: An Evolving Challenge in the Age of LLMs," 2023, *arXiv*. doi: 10.48550/arXiv.2309.15847.

[74] J. A. Leite, O. Razuvayevskaya, K. Bontcheva, and C. Scarton, "Detecting Misinformation with LLM-Predicted Credibility Signals and Weak Supervision," 2023, *arXiv*. doi: 10.48550/arXiv.2309.07601.

[75] J. Wu and B. Hooi, "Fake News in Sheep's Clothing: Robust Fake News Detection Against LLM-Empowered Style Attacks," 2023, *arXiv*. doi: 10.48550/arXiv.2310.10830.

[76] Anonymous, "The Earth is Flat because...: Investigating LLMs' Belief towards Misinformation via Persuasive Conversation," 2023, [Online]. Available: https://openreview.net/forum?id=DJXifFF2_M

[77] W. Guo and A. Caliskan, "Detecting Emergent Intersectional Biases: Contextualized Word Embeddings Contain a Distribution of Human-like Biases," in AIES '21. Association for Computing Machinery, 2021, pp. 122–133. doi: 10.1145/3461702.3462536.

[78] E. C. Choi and E. Ferrara, "Automated Claim Matching with Large Language Models: Empowering Fact-Checkers in the Fight Against Misinformation," Oct. 2023, [Online]. Available: http://arxiv.org/abs/2310.09223

[79] S. Lin, J. Hilton, and O. Evans, "TruthfulQA: Measuring How Models Mimic Human Falsehoods," 2022, *arXiv*. doi: 10.48550/arXiv.2109.07958.

[80] J. Lucas, A. Uchendu, M. Yamashita, J. Lee, S. Rohatgi, and D. Lee, "Fighting Fire with Fire: The Dual Role of LLMs in Crafting and Detecting Elusive Disinformation," 2023, *arXiv*. doi: 10.48550/arXiv.2310.15515.

[81] J. Su, T. Y. Zhuo, J. Mansurov, D. Wang, and P. Nakov, "Fake News Detectors are Biased against Texts Generated by Large Language Models," 2023, *arXiv*. doi: 10.48550/arXiv.2309.08674.

[82] K.-C. Yang and F. Menczer, "Large language models can rate news outlet credibility," 2023, *arXiv*. doi: 10.48550/arXiv.2304.00228.

[83] C. Chen and K. Shu, *Combating Misinformation in the Age of LLMs: Opportunities and Challenges*. 2023. [Online]. Available: files/9663/Chen and Shu - 2023 - Combating Misinformation in the Age of LLMs Oppor.pdf

[84] B. Xia, Q. Lu, L. Zhu, S. U. Lee, Y. Liu, and Z. Xing, "From Principles to Practice: An Accountability Metrics Catalogue for Managing AI Risks," 2023, *arXiv*. doi: 10.48550/arXiv.2311.13158.

[85] K. He *et al.*, "A Survey of Large Language Models for Healthcare: from Data, Technology, and Applications to Accountability and Ethics," 2023, *arXiv*. doi: 10.48550/arXiv.2310.05694.

[86] Y. Li, Y. Du, K. Zhou, J. Wang, W. X. Zhao, and J.-R. Wen, "Evaluating Object Hallucination in Large Vision-Language Models," 2023, *arXiv*. [Online]. Available: http://arxiv.org/abs/2305.10355

[87] S. A. Athaluri, S. V. Manthena, V. S. R. K. M. Kesapragada, V. Yarlagadda, T. Dave, and R. T. S. Duddumpudi, "Exploring the Boundaries of Reality: Investigating the Phenomenon of Artificial Intelligence Hallucination in Scientific Writing Through ChatGPT References," *Cureus*, vol. 15, no. 4, p. e37432, Dec. 2023, doi: 10.7759/cureus.37432.

[88] A. Deroy, K. Ghosh, and S. Ghosh, "How Ready are Pre-trained Abstractive Models and LLMs for Legal Case Judgement Summarization?," 2023, *arXiv*. doi: 10.48550/arXiv.2306.01248.

[89] A. Choudhury and O. Asan, "Impact of accountability, training, and human factors on the use of artificial intelligence in healthcare: Exploring the perceptions of healthcare practitioners in the us," *Human Factors in Healthcare vol. 2, p. 100021*, 2022.

[90] A. Rogers, "Changing the World by Changing the Data," *Proceedings of the 59th Annual Meeting of the Association for Computational Linguistics and the 11th International Joint Confer*, 2021, doi: 10.18653/v1/2021.acl-long.170.

[91] A. Birhane, "Algorithmic injustice: a relational ethics approach," *Patterns (N Y)*, vol. 2, no. 2, p. 100205, 2021, doi: 10.1016/j.patter.2021.100205.

[92] A. Paullada, I. D. Raji, E. M. Bender, E. Denton, and A. Hanna, "Data and its (dis)contents: A survey of dataset development and use in machine learning research," *Patterns (N Y)*, vol. 2, no. 11, p. 100336, 2021, doi: 10.1016/j.patter.2021.100336.

[93] M. D. J. V. and N. R. Anamaria Crisan, "Interactive Model Cards: A Human-Centered Approach to Model Documentation.," *In 2022 ACM Conference on Fairness, Accountability, and Transparency (FAccT '22). Association for Computing Machinery, New York, NY, USA, 427–439. https: //doi.org/10.1145/3531146.3533108* , 2022.

[94] A. Liesenfeld, A. Lopez, and M. Dingemanse, "Opening up ChatGPT: Tracking openness, transparency, and accountability in instruction-tuned text generators," in CUI '23. Association for Computing Machinery, 2023, pp. 1–6. doi: 10.1145/3571884.3604316.

[95] A. Liesenfeld, A. Lopez, and M. Dingemanse, "Opening up ChatGPT: Tracking openness, transparency, and accountability in instruction-tuned text generators," in CUI '23. Association for Computing Machinery, 2023, pp. 1–6. doi: 10.1145/3571884.3604316.

[96] A. Gudibande *et al.*, "The False Promise of Imitating Proprietary LLMs," 2023, *arXiv*. doi: 10.48550/arXiv.2305.15717.

[97] R. Mao, G. Chen, X. Zhang, F. Guerin, and E. Cambria, *GPTEval: A Survey on Assessments of ChatGPT and GPT-4*. 2023. [Online]. Available: files/10027/Mao et al. - 2023 - GPTEval A Survey on Assessments of ChatGPT and GP.pdf

[98] R. Mao, G. Chen, X. Zhang, F. Guerin, and E. Cambria, *GPTEval: A Survey on Assessments of ChatGPT and GPT-4*. 2023. [Online]. Available: files/10027/Mao et al. - 2023 - GPTEval A Survey on Assessments of ChatGPT and GP.pdf

[99] J. Huang and K. Chen-Chuan Chang, "Citation: A Key to Building Responsible and Accountable Large Language Models," 2023. doi: 10.48550/arXiv.2307.02185.

[100] E. Guo *et al.*, "neuroGPT-X: Towards an Accountable Expert Opinion Tool for Vestibular Schwannoma," vol. 1, 2023, doi: 10.17632/b9mck42r35.1.

[101] M. and H. A. Khan, "The Subjects and Stages of AI Dataset Development: A Framework for Dataset Accountability ," *Forthcoming 19 Ohio St. Tech. L.J. (2023), Available at SSRN: https://ssrn.com/abstract=4217148 or http://dx.doi.org/10.2139/ssrn.4217148*, 2022.

[102] G. E. Cacciamani *et al.*, "Development of the ChatGPT, Generative Artificial Intelligence and Natural Large Language Models for Accountable Reporting and Use (CANGARU) Guidelines," 2023, *arXiv*. [Online]. Available: http://arxiv.org/abs/2307.08974

[103] M. E. T. S. J. O. L. S. B. B. E. B. J. S. R. T. L. S. Anderljung, "Towards Publicly Accountable Frontier LLMs: Building an External Scrutiny Ecosystem under the ASPIRE Framework.," *arXiv preprint arXiv:2311.14711*, 2023.

[104] K. Nabben, "Constituting an AI: Accountability Lessons from an LLM Experiment," 2023, *Rochester, NY*. doi: 10.2139/ssrn.4561433.

[105] J. W. Allen, B. D. Earp, J. Koplin, and D. Wilkinson, "Consent-GPT: is it ethical to delegate procedural consent to conversational AI?," *J Med Ethics*, 2023, doi: 10.1136/jme-2023-109347.

[106] M. Jeyaraman, S. Balaji, N. Jeyaraman, and S. Yadav, "Unraveling the Ethical Enigma: Artificial Intelligence in Healthcare," *Cureus*, 2023, doi: 10.7759/cureus.43262.

[107] V. Rahimzadeh, K. Kostick-Quenet, J. Blumenthal Barby, and A. L. McGuire, "Ethics Education for Healthcare Professionals in the Era of ChatGPT and Other Large Language Models: Do We Still Need It?," *American Journal of Bioethics*, vol. 23, no. 10, pp. 17–27, 2023, doi: 10.1080/15265161.2023.2233358.

[108] H. Li, J. T. Moon, S. Purkayastha, L. A. Celi, H. Trivedi, and J. W. Gichoya, "Ethics of large language models in medicine and medical research," *Lancet Digit Health*, vol. 5, no. 6, pp. e333–e335, 2023, doi: 10.1016/S2589-7500(23)00083-3.

[109] K. He *et al.*, "A Survey of Large Language Models for Healthcare: from Data, Technology, and Applications to Accountability and Ethics," 2023, *arXiv*. doi: 10.48550/arXiv.2310.05694.

[110] C. Wang, S. Liu, H. Yang, J. Guo, Y. Wu, and J. Liu, "Ethical Considerations of Using ChatGPT in Health Care," *J Med Internet Res*, vol. 25, no. 1, p. e48009, 2023, doi: 10.2196/48009.

[111] A. Graf and R. E. Bernardi, “ChatGPT in Research: Balancing Ethics, Transparency and Advancement,” *Neuroscience*, vol. 515, pp. 71–73, 2023, doi: 10.1016/j.neuroscience.2023.02.008.

[112] B. D. Lund, T. Wang, N. R. Mannuru, B. Nie, S. Shimray, and Z. Wang, “ChatGPT and a new academic reality: Artificial Intelligence-written research papers and the ethics of the large language models in scholarly publishing,” *J Assoc Inf Sci Technol*, vol. 74, no. 5, pp. 570–581, 2023, doi: 10.1002/asi.24750.

[113] N. Dehouche, “Plagiarism in the age of massive Generative Pre-trained Transformers (GPT-3),” *Ethics Sci Environ Polit*, vol. 21, pp. 17–23, 2021, doi: 10.3354/esep00195.

[114] J. Y. Park, “Could ChatGPT help you to write your next scientific paper?: concerns on research ethics related to usage of artificial intelligence tools,” 2023, *Korean Association of Oral and Maxillofacial Surgeons*. doi: 10.5125/jkaoms.2023.49.3.105.

[115] L. A. Schintler, C. L. McNeely, and J. Witte, “A Critical Examination of the Ethics of AI-Mediated Peer Review,” 2023, *arXiv*. doi: 10.48550/arXiv.2309.12356.

[116] H. A. McKee and J. E. Porter, “Ethics for AI Writing: The Importance of Rhetorical Context,” in AIES ’20. Association for Computing Machinery, 2020, pp. 110–116. doi: 10.1145/3375627.3375811.

[117] N. F. Lindemann, “Sealed Knowledges: A Critical Approach to the Usage of LLMs as Search Engines,” in AIES ’23. Association for Computing Machinery, 2023, pp. 985–986. doi: 10.1145/3600211.3604737.

[118] D. H. R. Spennemann, “Exploring Ethical Boundaries: Can ChatGPT Be Prompted to Give Advice on How to Cheat in University Assignments?,” 2023, *Preprints*. doi: 10.20944/preprints202308.1271.v1.

[119] E. Kasneci *et al.*, “ChatGPT for good? On opportunities and challenges of large language models for education,” *Learn Individ Differ*, vol. 103, p. 102274, 2023, doi: 10.1016/j.lindif.2023.102274.

[120] S. Porsdam Mann, B. D. Earp, N. Møller, S. Vynn, and J. Savulescu, “AUTOGEN: A Personalized Large Language Model for Academic Enhancement—Ethics and Proof of Principle,” *The American Journal of Bioethics*, vol. 23, no. 10, pp. 28–41, 2023, doi: 10.1080/15265161.2023.2233356.

[121] D. Mhlanga, “Open AI in Education, the Responsible and Ethical Use of ChatGPT Towards Lifelong Learning,” 2023, *Rochester, NY*. doi: 10.2139/ssrn.4354422.

[122] U. Shakir, J. L. Hess, M. James, and A. Katz, “Pushing Ethics Assessment Forward in Engineering: NLP-Assisted Qualitative Coding of Student Responses,” in *2023 ASEE Annual Conference & Exposition*, 2023. [Online]. Available: https://peer.asee.org/pushing-ethics-assessment-forward-in-engineering-nlp-assisted-qualitative-coding-of-student-responses

[123] A. Basir, E. D. Puspitasari, C. C. Aristarini, P. D. Sulastri, and A. M. A. Ausat, “Ethical Use of ChatGPT in the Context of Leadership and Strategic Decisions,” *Jurnal Minfo Polgan*, vol. 12, no. 1, pp. 1239–1246, 2023, doi: 10.33395/jmp.v12i1.12693.

[124] M. Ryan, E. Christodoulou, J. Antoniou, and K. Iordanou, “An AI ethics ‘David and Goliath’: value conflicts between large tech companies and their employees,” *AI Soc*, 2022, doi: 10.1007/s00146-022-01430-1.

[125] N. Lourie, R. Le Bras, and Y. Choi, “SCRUPLES: A Corpus of Community Ethical Judgments on 32,000 Real-Life Anecdotes,” *Proceedings of the AAAI Conference on Artificial Intelligence*, vol. 35, no. 15, pp. 13470–13479, 2021, doi: 10.1609/aaai.v35i15.17589.

[126] S. Prabhumoye, B. Boldt, R. Salakhutdinov, and A. W. Black, “Case Study: Deontological Ethics in NLP,” 2021, *arXiv*. doi: 10.48550/arXiv.2010.04658.

[127] J. Cabrera, M. S. Loyola, I. Magaña, and R. Rojas, “Ethical Dilemmas, Mental Health, Artificial Intelligence, and LLM-Based Chatbots,” I. Rojas, O. Valenzuela, F. Rojas Ruiz, L. J. Herrera, and F. Ortuño, Eds., in Lecture Notes in Computer Science. Springer Nature Switzerland, 2023, pp. 313–326. doi: 10.1007/978-3-031-34960-7_22.

[128] C. Ashurst, E. Hine, P. Sedille, and A. Carlier, “AI Ethics Statements: Analysis and Lessons Learnt from NeurIPS Broader Impact Statements,” in FAccT ’22. Association for Computing Machinery, 2022, pp. 2047–2056. doi: 10.1145/3531146.3533780.

[129] A.-R. Bhojani and M. Schwarting, “Truth and Regret: Large Language Models, the Quran, and Misinformation,” *Theology and Science*, vol. 0, no. 0, pp. 1–7, 2023, doi: 10.1080/14746700.2023.2255944.

[130] Z. Su *et al.*, “InfoEntropy Loss to Mitigate Bias of Learning Difficulties for Generative Language Models,” 2023, doi: 10.48550/arXiv.2310.19531.

[131] P. Yu and H. Ji, “Self Information Update for Large Language Models through Mitigating Exposure Bias,” 2023, *arXiv*. doi: 10.48550/arXiv.2305.18582.

[132] J. S. Ernst *et al.*, “Bias Mitigation for Large Language Models using Adversarial Learning”, [Online]. Available: files/9597/Ernst et al. - Bias Mitigation for Large Language Models using Ad.pdf

[133] M. Xue *et al.*, “OccuQuest: Mitigating Occupational Bias for Inclusive Large Language Models,” 2023, doi: 10.48550/arXiv.2310.16517.

[134] Z. Zhang, L. Lyu, X. Ma, C. Wang, and X. Sun, "Fine-mixing: Mitigating Backdoors in Fine-tuned Language Models," 2022, doi: 10.48550/arXiv.2210.09545.

[135] J. Jang *et al.*, "Knowledge Unlearning for Mitigating Privacy Risks in Language Models," 2022, doi: 10.48550/arXiv.2210.01504.

[136] Z. He, H. Deng, H. Zhao, N. Liu, and M. Du, "Mitigating Shortcuts in Language Models with Soft Label Encoding," 2023, *arXiv*. doi: 10.48550/arXiv.2309.09380.

[137] H. Thakur, A. Jain, P. Vaddamanu, P. P. Liang, and L.-P. Morency, "Language Models Get a Gender Makeover: Mitigating Gender Bias with Few-Shot Data Interventions," 2023, *arXiv*. doi: 10.48550/arXiv.2306.04597.

[138] T. Dolci, "Fine-Tuning Language Models to Mitigate Gender Bias in Sentence Encoders," *2022 IEEE Eighth International Conference on Big Data Computing Service and Applications (BigDataService)*, pp. 175–176, 2022, doi: 10.1109/BigDataService55688.2022.00036.

[139] J. Zhao, M. Fang, Z. Shi, Y. Li, L. Chen, and M. Pechenizkiy, "CHBias: Bias Evaluation and Mitigation of Chinese Conversational Language Models," 2023, *arXiv*. doi: 10.48550/arXiv.2305.11262.

[140] A. Omrani *et al.*, "Social-Group-Agnostic Bias Mitigation via the Stereotype Content Model," in *Proceedings of the 61st Annual Meeting of the Association for Computational Linguistics (Volume 1: Long Papers)*, Association for Computational Linguistics, 2023, pp. 4123–4139. doi: 10.18653/v1/2023.acl-long.227.

[141] U. Gupta *et al.*, "Mitigating Gender Bias in Distilled Language Models via Counterfactual Role Reversal," 2022, *arXiv*. doi: 10.48550/arXiv.2203.12574.

[142] M. Bozdag, N. Sevim, and A. Koç, "Measuring and Mitigating Gender Bias in Legal Contextualized Language Models," *ACM Trans Knowl Discov Data*, 2023, doi: 10.1145/3628602.

[143] N. Varshney, W. Yao, H. Zhang, J. Chen, and D. Yu, "A Stitch in Time Saves Nine: Detecting and Mitigating Hallucinations of LLMs by Validating Low-Confidence Generation," 2023, doi: 10.48550/arXiv.2307.03987.

[144] H. Lee, S. Hong, J. Park, T. Kim, G. Kim, and J. Ha, "[Industry] KoSBI: A Dataset for Mitigating Social Bias Risks Towards Safer Large Language Model Applications," *Proceedings of the 61st Annual Meeting of the Association for Computational Linguistics*, 2023, [Online]. Available: https://virtual2023.aclweb.org/paper_I55.html

[145] X. Dong, Z. Zhu, Z. Wang, M. Teleki, and J. Caverlee, "Co$^2$PT: Mitigating Bias in Pre-trained Language Models through Counterfactual Contrastive Prompt Tuning," 2023, *arXiv*. doi: 10.48550/arXiv.2310.12490.

[146] D. Huang, Q. Bu, J. Zhang, X. Xie, J. Chen, and H. Cui, "Bias Assessment and Mitigation in LLM-based Code Generation," 2023, *arXiv*. doi: 10.48550/arXiv.2309.14345.

[147] R. Steed, S. Panda, A. Kobren, and M. Wick, "Upstream Mitigation Is *N*ot All You Need: Testing the Bias Transfer Hypothesis in Pre-Trained Language Models," in *ACL 2022*, S. Muresan, P. Nakov, and A. Villavicencio, Eds., Association for Computational Linguistics, 2022, pp. 3524–3542. doi: 10.18653/v1/2022.acl-long.247.

[148] H. Ngo *et al.*, "Mitigating harm in language models with conditional-likelihood filtration," 2021, doi: 10.48550/arXiv.2108.07790.

[149] S. Moon and N. Okazaki, "Effects and Mitigation of Out-of-vocabulary in Universal Language Models," *Journal of Information Processing*, vol. 29, pp. 490–503, 2021, doi: 10.2197/ipsjjip.29.490.

[150] J. Lu *et al.*, "Evaluation and Mitigation of Agnosia in Multimodal Large Language Models," 2023, *arXiv*. doi: 10.48550/arXiv.2309.04041.

[151] H. Van, "Mitigating Data Scarcity for Large Language Models," 2023, *arXiv*. doi: 10.48550/arXiv.2302.01806.

[152] H. Viswanath and T. Zhang, "FairPy: A Toolkit for Evaluation of Social Biases and their Mitigation in Large Language Models," 2023, *arXiv*. doi: 10.48550/arXiv.2302.05508.

[153] T. Shen, J. Li, M. R. Bouadjenek, Z. Mai, and S. Sanner, "Towards understanding and mitigating unintended biases in language model-driven conversational recommendation," *Inf Process Manag*, vol. 60, no. 1, p. 103139, 2023, doi: 10.1016/j.ipm.2022.103139.

[154] Z. Ji, T. Yu, Y. Xu, N. Lee, E. Ishii, and P. Fung, "Towards Mitigating Hallucination in Large Language Models via Self-Reflection," 2023, *arXiv*. doi: 10.48550/arXiv.2310.06271.

[155] F. Leiser, S. Eckhardt, M. Knaeble, A. Maedche, G. Schwabe, and A. Sunyaev, "From ChatGPT to FactGPT: A Participatory Design Study to Mitigate the Effects of Large Language Model Hallucinations on Users," pp. 81–90, 2023, doi: 10.1145/3603555.3603565.

[156] R. K. Mahabadi, Y. Belinkov, and J. Henderson, "End-to-End Bias Mitigation by Modelling Biases in Corpora," 2020, doi: 10.48550/arXiv.1909.06321.

[157] A. Garimella *et al.*, "He is very intelligent, she is very beautiful? On Mitigating Social Biases in Language Modelling and Generation," in *Findings 2021*, C. Zong, F. Xia, W. Li, and R. Navigli, Eds., Association for Computational Linguistics, 2021, pp. 4534–4545. doi: 10.18653/v1/2021.findings-acl.397.

[158] E. L. Ungless, A. Rafferty, H. Nag, and B. Ross, "A Robust Bias Mitigation Procedure Based on the Stereotype Content Model," 2022, doi: 10.48550/arXiv.2210.14552.

[159] K. Martin, *Ethics of Data and Analytics: Concepts and Cases*. CRC Press, 2022. [Online]. Available: https://books.google.com/books?id=E51kEAAAQBAJ

[160] Z. Liu, X. Zhang, and F. Peng, "Mitigating Unintended Memorization in Language Models Via Alternating Teaching," in *ICASSP 2023 - 2023 IEEE International Conference on Acoustics, Speech and Signal Processing (ICASSP)*, 2023, pp. 1–5. doi: 10.1109/ICASSP49357.2023.10096557.

[161] R. Liu, C. Jia, J. Wei, G. Xu, L. Wang, and S. Vosoughi, "Mitigating Political Bias in Language Models through Reinforced Calibration," *Proceedings of the AAAI Conference on Artificial Intelligence*, vol. 35, no. 17, pp. 14857–14866, 2021, doi: 10.1609/aaai.v35i17.17744.

[162] X. Jin, F. Barbieri, B. Kennedy, A. M. Davani, L. Neves, and X. Ren, "On Transferability of Bias Mitigation Effects in Language Model Fine-Tuning," 2021, *arXiv*. doi: 10.48550/arXiv.2010.12864.

[163] A. Graf and R. E. Bernardi, "ChatGPT in Research: Balancing Ethics, Transparency and Advancement," *Neuroscience*, vol. 515, pp. 71–73, 2023, doi: 10.1016/j.neuroscience.2023.02.008.

[164] Q. V. Liao and J. W. Vaughan, "AI Transparency in the Age of LLMs: A Human-Centered Research Roadmap," 2023, doi: 10.48550/arXiv.2306.01941.

[165] T. Wu, M. Terry, and C. J. Cai, "AI Chains: Transparent and Controllable Human-AI Interaction by Chaining Large Language Model Prompts," pp. 1–22, 2022, doi: 10.1145/3491102.3517582.

[166] N. Musacchio *et al.*, "Transparent machine learning suggests a key driver in the decision to start insulin therapy in individuals with type 2 diabetes," *J Diabetes*, vol. 15, no. 3, pp. 224–236, 2023, doi: 10.1111/1753-0407.13361.

[167] Z. Huang, S. Gutierrez, H. Kamana, and S. Macneil, "Memory Sandbox: Transparent and Interactive Memory Management for Conversational Agents," in UIST '23 Adjunct. Association for Computing Machinery, 2023, pp. 1–3. doi: 10.1145/3586182.3615796.

[168] J. Bang, B.-T. Lee, and P. Park, "Examination of Ethical Principles for LLM-Based Recommendations in Conversational AI," in *2023 International Conference on Platform Technology and Service (PlatCon)*, 2023, pp. 109–113. doi: 10.1109/PlatCon60102.2023.10255221.

[169] A. Liesenfeld, A. Lopez, and M. Dingemanse, "Opening up ChatGPT: Tracking openness, transparency, and accountability in instruction-tuned text generators," pp. 1–6, 2023, doi: 10.1145/3571884.3604316.

[170] D. Glukhov, I. Shumailov, Y. Gal, N. Papernot, and V. Papyan, "LLM Censorship: A Machine Learning Challenge or a Computer Security Problem?," 2023, doi: 10.48550/arXiv.2307.10719.

[171] A. Karamolegkou, J. Li, L. Zhou, and A. Søgaard, "Copyright Violations and Large Language Models," Oct. 2023, [Online]. Available: http://arxiv.org/abs/2310.13771

[172] N. Rahman and E. Santacana, "Beyond Fair Use: Legal Risk Evaluation for Training LLMs on Copyrighted Text", [Online]. Available: files/9726/Rahman and Santacana - Beyond Fair Use Legal Risk Evaluation for Trainin.pdf

[173] W. Peng *et al.*, "Are You Copying My Model? Protecting the Copyright of Large Language Models for EaaS via Backdoor Watermark," May 2023, [Online]. Available: http://arxiv.org/abs/2305.10036

[174] T. Chu, Z. Song, and C. Yang, "How to Protect Copyright Data in Optimization of Large Language Models?," Aug. 2023, [Online]. Available: http://arxiv.org/abs/2308.12247

[175] Y. Liu, H. Hu, X. Chen, X. Zhang, and L. Sun, "Watermarking Classification Dataset for Copyright Protection," 2023, *arXiv*. doi: 10.48550/arXiv.2305.13257.

[176] L. Waidelich, M. Lambert, Z. Al-Washash, S. Kroschwald, T. Schuster, and N. Döring, "Using Large Language Models for the Enforcement of Consumer Rights in Germany," J. Maślankowski, B. Marcinkowski, and P. Rupino da Cunha, Eds., in Lecture Notes in Business Information Processing. Springer Nature Switzerland, 2023, pp. 1–15. doi: 10.1007/978-3-031-43590-4_1.

[177] S. Kiritchenko, I. Nejadgholi, and K. C. Fraser, "Confronting Abusive Language Online: A Survey from the Ethical and Human Rights Perspective," *Journal of Artificial Intelligence Research*, vol. 71, pp. 431–478, 2021, doi: 10.1613/jair.1.12590.

[178] T. T. Nguyen, C. Wilson, and J. Dalins, "Fine-Tuning Llama 2 Large Language Models for Detecting Online Sexual Predatory Chats and Abusive Texts," 2023, doi: 10.48550/arXiv.2308.14683.

[179] D. Cortiz and A. Zubiaga, "Ethical and technical challenges of AI in tackling hate speech," *The International Review of Information Ethics*, vol. 29, 2020, doi: 10.29173/irie416.

[180] F. M. Plaza-del-arco, D. Nozza, and D. Hovy, “Respectful or Toxic? Using Zero-Shot Learning with Language Models to Detect Hate Speech,” in *WOAH 2023*, Y. Chung, P. R\textbackslash"ottger, D. Nozza, Z. Talat, and A. Mostafazadeh Davani, Eds., Association for Computational Linguistics, 2023, pp. 60–68. doi: 10.18653/v1/2023.woah-1.6.

[181] T. Hartvigsen, S. Gabriel, H. Palangi, M. Sap, D. Ray, and E. Kamar, “ToxiGen: A Large-Scale Machine-Generated Dataset for Adversarial and Implicit Hate Speech Detection,” 2022, doi: 10.48550/arXiv.2203.09509.

[182] V. K. Felkner, H.-C. H. Chang, E. Jang, and J. May, “WinoQueer: A Community-in-the-Loop Benchmark for Anti-LGBTQ+ Bias in Large Language Models,” 2023, doi: 10.48550/arXiv.2306.15087.

[183] D. Ottosson, “Cyberbullying Detection on social platforms using LargeLanguage Models,” 2023, [Online]. Available: https://urn.kb.se/resolve?urn=urn:nbn:se:miun:diva-48990

[184] J. Mökander, J. Schuett, H. R. Kirk, and L. Floridi, “Auditing large language models: a three-layered approach,” *AI and Ethics*, 2023, doi: 10.1007/s43681-023-00289-2.

[185] F. Mireshghallah, “Auditing and Mitigating Safety Risks in Large Language Models,” 2023. [Online]. Available: https://escholarship.org/uc/item/28f9b6px

[186] Y. Zhang, B. Fitzgibbon, D. Garofolo, A. Kota, E. Papenhausen, and K. Mueller, “An Explainable AI Approach to Large Language Model Assisted Causal Model Auditing and Development”, [Online]. Available: files/9681/Zhang et al. - An Explainable AI Approach to Large Language Model.pdf

[187] E. Jones, A. Dragan, A. Raghunathan, and J. Steinhardt, “Automatically Auditing Large Language Models via Discrete Optimization,” 2023, *arXiv*. doi: 10.48550/arXiv.2303.04381.

[188] H. Hasanbeig, H. Sharma, L. Betthauser, F. V. Frujeri, and I. Momennejad, “ALLURE: Auditing and Improving LLM-based Evaluation of Text using Iterative In-Context-Learning,” 2023, *arXiv*. doi: 10.48550/arXiv.2309.13701.

[189] T. L. Föhr, K.-U. Marten, and M. Schreyer, “Deep Learning Meets Risk-Based Auditing: a Holistic Framework for Leveraging Foundation and Task-Specific Models in Audit Procedures,” 2023, *Rochester, NY*. doi: 10.2139/ssrn.4488271.

[190] C. Rastogi, M. Tulio Ribeiro, N. King, H. Nori, and S. Amershi, “Supporting Human-AI Collaboration in Auditing LLMs with LLMs,” in AIES ’23. Association for Computing Machinery, 2023, pp. 913–926. doi: 10.1145/3600211.3604712.

[191] C. Wei, Y.-C. Wang, B. Wang, and C.-C. J. Kuo, “An Overview on Language Models: Recent Developments and Outlook,” 2023, *arXiv*. doi: 10.48550/arXiv.2303.05759.

[192] C. Wei, Y.-C. Wang, B. Wang, and C.-C. J. Kuo, “An Overview on Language Models: Recent Developments and Outlook,” 2023, *arXiv*. doi: 10.48550/arXiv.2303.05759.

[193] C. Wei, Y.-C. Wang, B. Wang, and C.-C. J. Kuo, “An Overview on Language Models: Recent Developments and Outlook,” 2023, *arXiv*. doi: 10.48550/arXiv.2303.05759.

[193] Kokhlikyan, N., Miglani, V., Martin, M., Wang, E., Alsallakh, B., Reynolds, J., ... & Reblitz-Richardson, O. (2020). Captum: A unified and generic model interpretability library for pytorch. arXiv preprint arXiv:2009.07896.

[194] Miglani, V., Yang, A., Markosyan, A. H., Garcia-Olano, D., & Kokhlikyan, N. (2023). Using captum to explain generative language models. arXiv preprint arXiv:2312.05491.

[195] Tufanov, I., Hambardzumyan, K., Ferrando, J., & Voita, E. (2024). LM transparency tool: Interactive tool for analyzing transformer language models. arXiv preprint arXiv:2404.07004.

[196] Nadeem, M., Bethke, A., & Reddy, S. (2020). StereoSet: Measuring stereotypical bias in pretrained language models. arXiv preprint arXiv:2004.09456.

[197] Nangia, N., Vania, C., Bhalerao, R., & Bowman, S. R. (2020). CrowS-pairs: A challenge dataset for measuring social biases in masked language models. arXiv preprint arXiv:2010.00133.

[198] Wang, S., Li, R., Chen, X., Yuan, Y., Wong, D. F., & Yang, M. (2025). Exploring the impact of personality traits on llm bias and toxicity. arXiv preprint arXiv:2502.12566.

[199] Puttick, A., Rankwiler, L., Ikae, C., & Kurpicz-Briki, M. (2024). The BIAS Detection Framework: Bias Detection in Word Embeddings and Language Models for European Languages. arXiv preprint arXiv:2407.18689.

[200] May, C., Wang, A., Bordia, S., Bowman, S. R., & Rudinger, R. (2019). On measuring social biases in sentence encoders. arXiv preprint arXiv:1903.10561.

[201] Binkyte, R. (2025). Interactional Fairness in LLM Multi-Agent Systems: An Evaluation Framework. arXiv preprint arXiv:2505.12001.

[202] Achiam, J., Adler, S., Agarwal, S., Ahmad, L., Akkaya, I., Aleman, F. L., ... & McGrew, B. (2023). Gpt-4 technical report. arXiv preprint arXiv:2303.08774.
[203] Anthropic. (2024). The Claude 3 Model Family: Opus, Sonnet, Haiku
[204] Team, G., Anil, R., Borgeaud, S., Alayrac, J. B., Yu, J., Soricut, R., ... & Blanco, L. (2023). Gemini: a family of highly capable multimodal models. arXiv preprint arXiv:2312.11805.
[205] Conneau, A., Lample, G., Rinott, R., Williams, A., Bowman, S. R., Schwenk, H., & Stoyanov, V. (2018). XNLI: Evaluating cross-lingual sentence representations. arXiv preprint arXiv:1809.05053.
[206] Goyal, N., Gao, C., Chaudhary, V., Chen, P. J., Wenzek, G., Ju, D., ... & Fan, A. (2022). The flores-101 evaluation benchmark for low-resource and multilingual machine translation. Transactions of the Association for Computational Linguistics, 10, 522-538.
[207] Davani, A. M., Atari, M., Kennedy, B., & Dehghani, M. (2023). Hate speech classifiers learn normative social stereotypes. Transactions of the Association for Computational Linguistics, 11, 300-319.
[208] Kennedy, B., Golazizian, P., Trager, J., Atari, M., Hoover, J., Mostafazadeh Davani, A., & Dehghani, M. (2023). The (moral) language of hate. PNAS nexus, 2(7), pgad210.
[209] Omrani Sabbaghi, S., Wolfe, R., & Caliskan, A. (2023, August). Evaluating biased attitude associations of language models in an intersectional context. In Proceedings of the 2023 AAAI/ACM Conference on AI, Ethics, and Society (pp. 542-553).
[210] Nicolas, G., & Caliskan, A. (2024). A taxonomy of stereotype content in large language models. arXiv preprint arXiv:2408.00162.
[211] Nicolas, G., & Caliskan, A. (2024). Directionality and representativeness are differentiable components of stereotypes in large language models. PNAS nexus, 3(11), pgae493.
[212] Omrani, A., Salkhordeh_Ziabari, A., Yu, C., Golazizian, P., Kennedy, B., Atari, M., ... & Dehghani, M. (2023, January). Social-group-agnostic bias mitigation via the stereotype content model. Association for Computational Linguistics.
[213] Xiao, Y., Liu, A., Liang, S., Liu, X., & Tao, D. (2025). Fairness Mediator: Neutralize Stereotype Associations to Mitigate Bias in Large Language Models. arXiv preprint arXiv:2504.07787.